\documentclass[trackchanges, twocolumn]{aastex701}
\usepackage{url}
\usepackage{amsmath}

\usepackage{fmtcount}
\shorttitle{$M_\mathrm{BH}$--$M_\mathrm{bulge}$ Evolution from the GWB}
\shortauthors{Matt et al.}

\begin{document}

\title{Inferring $M_\mathrm{BH}$--$M_\mathrm{bulge}$ Evolution from the Gravitational Wave Background}

\correspondingauthor{Cayenne Matt}

\email[show]{cayenne@umich.edu}

\author{Cayenne Matt \orcid{0000-0002-9710-6527}}
\affiliation{Department of Astronomy and Astrophysics, University of Michigan, Ann Arbor, MI 48109, USA}
\email{cayenne.matt@nanograv.org}
\author{Kayhan G\"{u}ltekin \orcid{0000-0002-1146-0198}}
\affiliation{Department of Astronomy and Astrophysics, University of Michigan, Ann Arbor, MI 48109, USA}
\email{kayhan@umich.edu}
\author{Luke Zoltan Kelley \orcid{0000-0002-6625-6450}}
\affiliation{Department of Astronomy, University of California, Berkeley, 501 Campbell Hall \#3411, Berkeley, CA 94720, USA}
\email{luke.kelley@nanograv.org}
\author{Laura Blecha \orcid{0000-0002-2183-1087}}
\affiliation{Physics Department, University of Florida, Gainesville, FL 32611, USA}
\email{laura.blecha@nanograv.org}
\author{Joseph Simon \orcid{0000-0003-1407-6607}}
\altaffiliation{NSF Astronomy and Astrophysics Postdoctoral Fellow}
\affiliation{Department of Astrophysical and Planetary Sciences, University of Colorado, Boulder, CO 80309, USA}
\email{joe.simon@nanograv.org}
\author{Gabriella Agazie \orcid{0000-0001-5134-3925}}
\affiliation{Center for Gravitation, Cosmology and Astrophysics, Department of Physics, University of Wisconsin-Milwaukee,\\ P.O. Box 413, Milwaukee, WI 53201, USA}
\email{gabriella.agazie@nanograv.org}
\author{Akash Anumarlapudi \orcid{0000-0002-8935-9882}}
\affiliation{Department of Physics and Astronomy, University of North Carolina, Chapel Hill, NC 27599, USA}
\email{akasha@unc.edu}
\author{Anne M. Archibald \orcid{0000-0003-0638-3340}}
\affiliation{Newcastle University, NE1 7RU, UK}
\email{anne.archibald@nanograv.org}
\author{Zaven Arzoumanian \orcid{0009-0008-6187-8753}}
\affiliation{X-Ray Astrophysics Laboratory, NASA Goddard Space Flight Center, Code 662, Greenbelt, MD 20771, USA}
\email{zaven.arzoumanian@nanograv.org}
\author{Jeremy G. Baier \orcid{0000-0002-4972-1525}}
\affiliation{Department of Physics, Oregon State University, Corvallis, OR 97331, USA}
\email{jeremy.baier@nanograv.org}
% ...





\newcommand{\msigma}{$M_\mathrm{BH}$--$\sigma$}

\newcommand{\mmb}{$M_\mathrm{BH}$--$M_\mathrm{bulge}$}

\newcommand{\modelnum}{8}

\newcommand{\asa}{Le11ne}
\newcommand{\asaev}{Le11ev}

\newcommand{\asm}{Le03ne}
\newcommand{\asmev}{Le03ev}

\newcommand{\asb}{Le00ne}
\newcommand{\asbev}{Le00ev}

\newcommand{\liep}{LM11ne}
\newcommand{\liepev}{LM11ev}

\newcommand{\resalpha}{1.04}
\newcommand{\restd}{0.5}

\newcommand{\resalphaall}{0.94}
\newcommand{\restdall}{0.7}

\newcommand{\fnum}{5}

\newcommand{\orcid}[1]{\href{https://orcid.org/#1}{\includegraphics[scale=0.04]{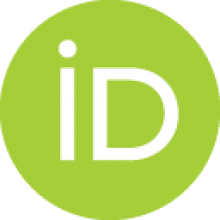}}}

\newcommand{\rev}{}
\newcommand{\revii}{}

\begin{abstract}

We test the impact of an evolving supermassive black hole (SMBH) mass scaling relation (\mmb) on the predictions for the gravitational wave background (GWB). The observed GWB amplitude is 2-3 times higher than predicted by astrophysically informed models which suggests the need to revise the assumptions in those models. We compare a semi-analytic model's ability to reproduce the observed GWB spectrum with a static versus evolving-amplitude \mmb\ relation. We additionally consider the influence of the choice of galaxy stellar mass function on the modeled GWB spectra. Our models are able to reproduce the GWB amplitude with either a large number density of massive galaxies or a positively evolving \mmb\ amplitude (i.e., the $M_\mathrm{BH} / M_\mathrm{bulge}$ ratio was higher in the past). If we assume that the \mmb\ amplitude does not evolve, our models require a galaxy stellar mass function that implies an undetected population of massive galaxies ($M_\star \geq 10^{11} M_\odot$ at $z > 1$). When the \mmb\ amplitude is allowed to evolve, we can model the GWB spectrum with all fiducial values and an \mmb\ amplitude that evolves as $\alpha(z) = \alpha_0 (1 + z)^{\resalpha \pm \restd}$.

\end{abstract}



\section{Introduction}\label{sec:intro}

In mid-2023, four pulsar timing array (PTA) collaborations announced evidence for the gravitational wave background (GWB) \citep{Hellings_1983, Agazie_2023, Antoniadis_2023, Reardon_2023, Xu_2023}. In all four announcements, the strain amplitude of the signal was 2--3 times greater than expected. It is commonly thought that the source for this quadrupolar signal is dominated by supermassive black hole (SMBH) binaries \citep{Begelman_1980, Milosavljevic_2001, Burke_Spolaor_2019}. In this case, the GWB amplitude is most sensitive to the chirp mass of the binaries and is therefore expected to be dominated by the most massive binaries \citep{Phinney_2001}. Thus, to model the GWB, it is necessary to have a reliable way to characterize the mass distribution of the underlying SMBH binary population. To make meaningful astrophysical inferences from PTA data it is therefore important to have an accurate predictor for the black hole mass function (BHMF), especially for the most massive SMBHs.

In the local universe, it has been shown that there is a strong correlation between host galaxy bulge mass and central SMBH mass, making it possible to accurately predict SMBH masses in nearby galaxies \rev{\citep{Kormendy_Ho_2013, McConnell_2013}}. Direct, independent SMBH mass measurements are difficult outside the local universe, however, and so very few constraints exist for this relation for $z > 0$. High redshift SMBH mass measurements are hindered by telescope resolution limits meaning that masses must be inferred indirectly. Because of their extreme luminosity, AGN can be used to place useful constraints on SMBH masses at redshifts as high as $z \sim 10$ \citep{Jeon_2025, Bogdan_2024, Maiolino_411_pop_2024, Napolitano_2024}. From the AGN luminosity function, one can infer the BHMF, but assumptions about, e.g., radiative efficiency and AGN fraction lead towards large uncertainties \citep{Shen_2020}. Furthermore, these types of surveys are frequently magnitude limited and are thus subject to observational biases \citep{Lauer_2007}.

While AGN studies inform SMBH populations, modeling the GWB calls for a simpler BHMF framework. The observed correlation between galaxy bulge mass and SMBH mass is known as the \mmb\ relation \citep{Kormendy_Ho_2013, McConnell_2013}. To approximate SMBH number density at higher redshifts, one can convolve the local \mmb\ relation with the galaxy stellar mass function (GSMF), which is well-observed for the redshifts where most of the GWB signal is expected to be produced \citep[$0 < z < 2$][]{Sesana_2004, AgazieBHB_2023, Leja_2020}. This method assumes the \mmb\ relation to be unchanging with time, an assumption with ambiguous observational support. \rev{It expected that analyses of the GWB can be used to place constraints on the underlying population of galaxies and black holes \citep[e.g.,][]{Simon_2016, Izquierdo-Villalba_2022, Bonoli_2025}.} \citet{AgazieBHB_2023} derived physical quantities from fitting the NANOGrav 15-year PTA data and their results suggested that the data are best described with a number density of SMBHs, with $M_\mathrm{BH} \gtrsim 10^9 M_\odot$, that is notably greater than predictions from current models \citep[such as ][]{Kormendy_Ho_2013, McConnell_2013}. Since then, other studies have used a variety of techniques finding similar results \citep[e.g.,][]{Sato-Polito_2024, Chen_2024, Liepold_2024}. Additionally, there exist other scaling relations, such as the \msigma\ relation, which connects SMBH mass to galaxy velocity dispersion \citep{Ferrarese_2000, Gebhardt_2000}. It has been shown that the \mmb\ and \msigma\ relations make different predictions for the BHMF at high redshift such that the predicted GWB is higher when using \msigma\ \citep{Matt_2023, Simon_2023}. This difference in BHMF predictions may be due to evolution in one or both relations that is not accounted for. Via the three-way relationship between galaxy stellar mass, radius, and velocity dispersion \citep{de_Graaff_2021}, the \msigma\ relation naturally accounts for galaxy downsizing \citep{van_der_Wel_2014} and may therefore be a more fundamental probe of the SMBH's gravitational potential well than \mmb\ \citep{van_den_Bosch_2016, Cohn_2025}. This combination of findings suggests that methods of predicting SMBH number density are in need of revision.

It is unknown when the $z = 0$ \mmb\ relation was first established or how accurate it is outside the local universe. The strong local correlation between SMBH mass and galaxy bulge mass suggests the growth of these objects is coupled \citep{Kormendy_Ho_2013}. If the \mmb\ relation were to be unchanging with time, it would suggest a lockstep growth for galaxies and SMBHs. In other words, galaxy star formation matches the pace of SMBH accretion and / or the dominant growth mechanisms for both of these bodies are through major mergers or positive AGN/stellar feedback. It is unknown, however, whether SMBHs grow faster than their host galaxies, vice versa, or if they grow symbiotically over cosmic time. Whether a galaxy's SMBH is over- or undermassive relative to the \mmb\ relation throughout time has implications for the interactions between feedback from AGN accretion and galactic star formation \citep{Zhuang_2023, Cohn_2025}. Constraints on these growth pathways are necessary for determining SMBH number density and galaxy evolution outside the local universe.

Results from the \textit{James Webb Space Telescope} (JWST) have found galaxies at high redshift ($z > 4$) with overmassive black holes entirely unpredicted by the local \mmb\ relation \citep[e.g, ][]{Harikane_2023, Pacucci_2023, Matthee_2024, Juodzbalis_2025}. These high-redshift black holes are unexpectedly high in both mass and number density, with candidates up to $z \sim 11$ \citep[e.g.][]{Maiolino_single_11_2024}. Furthermore, there is an abundance of observational and simulation-based studies that have found evidence both for and against an evolution in the amplitude of the \mmb\ relation \citep[e.g.,][]{Wyithe_2003, Cattaneo_2005, Peng_2006, Croton_2006, Hopkins_2009, Jahnke_2009, Decarli_2010, Merloni_2010, Trakhtenbrot_2010, Bennert_2011, Cisternas_2011, Ding_2020, Li_2021, Zhang_2023, Pacucci_2023, Terrazas_2024, Yue_2024, Chen_2024, Cloonan_2024, Sah_2024a, Sah_2024b, Shimizu_2024, Pacucci_2024, Hoshi_2024, Kozhikkal_2024, Sun_2025, Tanaka_2025}. 

An \mmb\ amplitude that changes with cosmic time would have a significant impact on GWB predictions and interpretations. For example, if SMBH growth generally outpaces galaxy growth at higher redshifts, the $z = 0$ \mmb\ relation would therefore underestimate SMBH masses which would, in turn, lead to an underestimate of the GWB amplitude. Past studies have used electromagnetic (EM) observation, theory, and simulations to constrain potential evolution in the \mmb\ relation, and now gravitational waves offer a new, independent, basis for testing.

If, instead, the \mmb\ relation that is measured locally has not changed significantly since $z \sim 3$, a higher number density of massive galaxies in this redshift range could explain the high GWB amplitude. The galaxy stellar mass function (GSMF) is not observed to undergo significant evolution from $z = 1$ to $z = 0$ \citep[e.g.,][]{Leja_2020} though theory predicts otherwise \citep[e.g.,][]{Tacchella_2019}. Recently \citet{Liepold_2024} found that this lack of observed evolution may be an artifact of the survey design. They explain that, locally, the most massive galaxies are rare enough that most current integral-field spectroscopic surveys do not observe a large enough volume to catch them. This leads to an incomplete local sample for galaxies with masses $M_\star \geq 10^{11.5} M_\odot$. Using the MASSIVE survey \citep[which has a $107\arcsec \times 107\arcsec$ field of view, ][]{Ma_2014} \citet{Liepold_2024} measured a new $z = 0$ GSMF based on this sample and found that the number density of these massive galaxies is significantly higher than other measurements \citep[such as ][]{Bernardi_2013, Moustakas_2013, DSouza_2015, Leja_2020}. Their work suggests that predictions for the GWB may have previously been under counting the number of the most massive galaxies, and therefore SMBHs, which could lead to an underestimate of the GWB amplitude. This intriguing possibility warrants further testing.

In this paper we evaluate the possibility of an evolving \mmb\ amplitude and test the impact of this evolution on the GWB spectrum. We additionally consider changes to the GSMF and explore the degeneracy between these two solutions. In section \ref{sec:methods} we describe our model setup including our functional form of redshift evolution of the \mmb\ relation. In section \ref{sec:results} we present the results of our models. We discuss the implications of our work in section \ref{sec:discussion}, and a summary of our work and conclusions can be found in section \ref{sec:summary_and_conclusions}.

\section{Methods}\label{sec:methods}

In this section we describe the general setup of our semi-analytic model and our choices for the \mmb\ scaling relation and GSMF. We additionally provide details for the priors and assumptions for our different test cases in section \ref{sec:the_models}.

\subsection{Semi-Analytic Modeling}\label{sec:holodeck}
\setcounter{footnote}{0}
We use \textsc{holodeck} \citep[details can be found in section 3 of][]{AgazieBHB_2023}\footnote{Also see \url{https://github.com/nanograv/holodeck}.} to synthesize populations of massive black holes. We start by convolving the galaxy stellar mass function from \citet{Leja_2020} with the SMBH--galaxy mass scaling relation from \citep{Kormendy_Ho_2013}. The specifics of this implementation are detailed in sections \ref{sec:methods_mmb_ev} and \ref{sec:methods_gsmf}. We then apply galaxy merger rates, and SMBH binary hardening models to solve for the number density of SMBH binaries emitting gravitational waves in frequencies detectable by PTAs, i.e., the GWB spectrum predicted from the input parameters. Previously, \citet{AgazieBHB_2023} calculated galaxy merger rates from galaxy pair fractions and merger times following the prescription due to \citep{Chen_PF_2019}. For this work, we instead use the galaxy merger rate prescription due to \citet{Rodriguez_Gomez_2015}\rev{, which provides a more astrophysically motivated model for mergers. This change does not significantly affect the shape or amplitude of the GWB spectrum (see Figure \ref{fig:kh_vs_mcma} in Appendix \ref{sec:app_gsmf}).} A detailed description and graphic of the \textsc{holodeck} workflow can be found in section 3 of \citet{AgazieBHB_2023}.

To determine our best-fit parameters for each model, we first sampled 20,000 times from a set of astrophysically-motivated prior distributions with a Latin hypercube. We generated and averaged over 100 realizations for each set of sampled parameters to account for the Poisson sampling in the GWB spectrum calculations. Our sample size is higher than was used in \citet{AgazieBHB_2023}, who used Gaussian process interpolation for their posterior parameter estimation. Because of our larger parameter space (up to 30 parameters versus their 6), it is not computationally feasible to follow their same methods \citep{Laal_2025}. Instead, our finer sampling makes it so that we can compare each model to the spectrum without the need for Gaussian process interpolation between the models. \rev{We compare our models to the 15 year Hellings and Downs (HD)-correlated free-spectrum representation of the GWB \citep{Lamb_2023}. Specifically, we use the ``HD-w/MP+DP+CURN" model which included both monopole (MP) and dipole (DP) correlated red noise in addition to the common uncorrelated red noise (CURN) \citep[see additional details in][]{AgazieBHB_2023}.} For each GWB frequency, we evaluate the goodness of fit for a given modeled spectrum by comparing the value of each model to the probability density distribution of the data to determine how well each model fits the data. This process produces results with the same fidelity as \citet{AgazieBHB_2023} with a higher computational efficiency (see their Appendix C, also our figures in Appendix \ref{sec:app_comp}).

\revii{Throughout this paper we refer to the ``likelihood'' of the models which we calculate in our fitting process. These likelihoods represent how well models fit the GWB data, but do not contain information on how well the models agree with observational constraints from EM datasets. Therefore we additionally quantify goodness of fit to, e.g. the GSMF, with statistical tests such as the Kullback-Leilbler divergence \citep[$D_{KL}$, ][see Section \ref{sec:results}]{Kullback_1951}, and a quantity, $\Xi$, which we define in Section \ref{sec:res_gsmf_choice} to evaluate the agreement between our models and the observed GSMF.}

\subsection{The \texorpdfstring{\mmb}{black hole mass bulge mass} relation}\label{sec:methods_mmb_ev}

The local relation between SMBH and galaxy bulge mass can be described by a power-law relation \citep[see, e.g.,][]{Kormendy_2001, Kormendy_Ho_2013, McConnell_2013}. The value of the $y$-intercept (amplitude) of this relation, throughout time, encodes the extent to which SMBHs and galaxies grow at the same rate. A constant amplitude value means that growth is tightly coupled and the coupling mechanisms are constant throughout cosmic time. In this work, we model a changing amplitude with a power-law evolution. We modified the existing \mmb\ framework in \textsc{holodeck} to allow for this evolution by replacing the constant amplitude with one that is a function of redshift, $\alpha_0 \rightarrow \alpha(z)$. We parameterize it as
\begin{equation}
    M_\mathrm{BH}=\alpha(z) \left(\frac{M_{\mathrm {bulge}}}{10^{11}\ M_{\odot}}\right)^{\beta_0}
	\label{eqn:mm}
\end{equation}
with
\begin{equation}
    \alpha(z) = \alpha_0 (1 + z)^{\alpha_z},
	\label{eqn:alphaz}
\end{equation}
where the $z = 0$ values can be determined from observation \citep[$\log \alpha_0 = 8.69$; $\beta_0 = 1.17$; ][]{Kormendy_Ho_2013}, and $\alpha_z$ can be positive or negative with $\alpha_z = 0$ indicating no evolution in the relation.

This power-law form of evolution is more rapid at lower redshifts, which means that the changes in the \mmb\ relation are greatest in the redshift range most relevant to the GWB \citep[$0 < z < 2$, ][]{Sesana_2004, AgazieBHB_2023}. We tested several functional forms of this evolution and determined that other options either evolve too slowly in the relevant redshift range, or are not distinguishable from a power-law with the current data. Future studies with higher signal-to-noise PTA data may eventually be able to place constraints on different functional forms, but until then, the power-law form is sufficient for this analysis.
Throughout this analysis, we use the \citet{Kormendy_Ho_2013} values for the local \mmb\ relation; we briefly discuss the impact of using alternative fits in Appendix \ref{sec:app_gsmf}.

\subsection{The Galaxy Stellar Mass Function}\label{sec:methods_gsmf}

\citet{AgazieBHB_2023} used the single-Schechter \citep{Schechter_1976} GSMF prescription due to \citet{Chen_2019}. For this work, we predominantly use the GSMF due to \citet{Leja_2020}, which is defined as a double-Schechter function offering a more accurate description of the mass function of the total galaxy population \citep[e.g.,][and references therein]{McLeod_2021}. This GSMF is strongly supported by observational data and has an explicitly defined evolution making it possible to reliably calculate the GSMF at any specific redshift. \rev{This model is reliable within the range of the data ($0.2 < z < 3$) though we extrapolate the model to higher redshifts in \textsc{holodeck} for defining our populations. The reliability of extrapolating this model is uncertain, but the majority of the GWB signal is dominated by lower redshifts \citep{AgazieBHB_2023, Sesana_2004} so we do not expect our results to be influenced by any error that may be introduced by this extrapolation.} \citet{Leja_2020} find that the redshift evolution of the characteristic mass and each of the density normalization terms is best described by a Gaussian. Following their methods (see their equations 14 and B 1-4), we define the full functional form of the GSMF to be
\begin{equation}
\begin{split}
\Phi(M_\star, z)  = \ln(10) \, \exp\left[ -\frac{M_\star}{M_\mathrm{c}(z)} \right] \times \\
\Bigg( \phi_{*, 1}(z) \left( \frac{M_\star}{M_\mathrm{c}(z)} \right)^{\alpha_1+1} 
+ \, \phi_{*, 2}(z) \left( \frac{M_\star}{M_\mathrm{c}(z)} \right)^{\alpha_2+1} \Bigg)
\end{split}
\label{eqn:double_schechter}
\end{equation}
where
\begin{equation}
    \log \phi_{*, i}(z) = {\phi_{*, i,0} + \phi_{*, i,1}  z + \phi_{*, i,2}  z^2},
	\label{eqn:phi}
\end{equation}
and
\begin{equation}
    \log M_\mathrm{c}(z) =  {M_{\mathrm{c} ,0} + M_{\mathrm{c} ,1}  z + M_{\mathrm{c} ,2}  z^2}.
	\label{eqn:mchar}
\end{equation}
$M_\mathrm{c}(z)$ is the characteristic mass, $\phi_{*, i}(z)$ is density normalization, and $\alpha_1$ and $\alpha_2$ are the upper and lower slopes of the power-law.

Recently, \citet{Liepold_2024} developed a $z = 0$ GSMF which has a notably higher number density than that of \citet{Leja_2020} for $M_\star \geq 10^{11.5} M_\odot$. They make an estimate of the GWB amplitude and conclude that, using their local GSMF and an implied redshift evolution informed by the fractional GWB contribution functions from \citet{AgazieBHB_2023} (see their Figure 12). Using their GSMF they claim they can produce a GWB amplitude that is consistent with what is seen by PTAs. Motivated by their results, we also consider how changes to the GSMF could instead explain the discrepancy between the predicted and observed GWB amplitude. \citet{Liepold_2024} do not provide an explicit evolutionary form for their GSMF and so we assume an evolution that is equivalent to their $z = 0$ GSMF and consistent with \citet{Leja_2020} for $z \gtrsim 1$. Further details of this evolution and discussion of alternative models can be found in Appendix \ref{sec:app_gsmf}. 

\subsection{The Models}\label{sec:the_models}

We present \modelnum\ models, which are summarized in Table \ref{tab:models}. There are four different parameter setups and for each setup we ran one version of the model using the non-evolving \mmb\ relation, and one version with evolution. For the models that allow for evolution of the \mmb\ amplitude, we used a uniform prior for the evolutionary parameter $-3 \leq \alpha_z \leq 3$.

The largest models include 29 parameters (30 with $\alpha_z$). Of these parameters, the GSMFs make up 11 variables, five that describe the local GSMF and six that contain information about the GSMF evolution. \rev{For six of our \numberstringnum{\modelnum} models the priors for sampled GSMF parameters are based on the posterior fit values from \citet{Leja_2020}. Our fiducial parameters are therefore the median value of these distributions. For two of our \numberstringnum{\modelnum} we use the same functional form of the GSMF, but with local values due to \citet{Liepold_2024} and evolutionary parameters we define for this work (see appendix \ref{sec:app_gsmf}).} To help understand the degeneracy between a changing GSMF versus \mmb\ relation we consider three different cases of sampling the \rev{\citet{Leja_2020}} GSMF: one in which we sample all 11 variables (\asa\ and \asaev), one in which we only sample three local values (\asm\ and \asmev), and one with all GSMF parameters fixed to their fiducial values (\asb\, \asbev). The three GSMF values that are varied in \asm\ and \asmev\ are $\log \phi_{*, 1, 0}$, $\log \phi_{*, 2, 0}$, and $\log M_{c, 0}$. For all but two of our \modelnum\ models we use the Gaussian posteriors from \citet{Leja_2020} as priors for our GSMF and we assume that these parameters are independent of each other. For the last two models, we use our explicitly evolving version of the \citet{Liepold_2024} GSMF, which we described previously in section \ref{sec:methods_gsmf} and also in appendix section \ref{sec:app_gsmf}.

Each model we present is given a short-hand name to indicate the important features. Each name is six total characters to convey three pieces of information: The first two characters indicate which version of the GSMF is being used where Le is \citet{Leja_2020} and LM is \citet{Liepold_2024}. The middle two characters indicate how many of the GSMF parameters (out of 11) were sampled where the number of parameters fixed to their fiducial values is 11 minus that number. The final two characters indicate whether the \mmb\ amplitude was allowed to evolve (ev) or not (ne). This information is summarized in Table \ref{tab:models}.

\begin{deluxetable}{llc}
\tablewidth{0pt}
\tablehead{
\colhead{Model Name} & \colhead{GSMF Parameters} & \colhead{Evolving \mmb}
}
\startdata
\asa\ & 11 sampled & No \\
\asaev\ & 11 sampled & Yes \\
\asm\ & 3 sampled & No \\
\asmev\ & 3 sampled & Yes \\
\asb\ & 0 sampled & No \\
\asbev\ & 0 sampled & Yes \\
\liep\ & 11  sampled & No \\
\liepev\ & 11  sampled & Yes \\
\enddata
\caption{A summary of the parameter set up for the \modelnum\ models in this paper. The naming convention for our models is xxyyzz where xx gives the source for the GSMF (Le for \citet{Leja_2020} and LM for \citet{Liepold_2024}), yy dictates the number of GSMF parameters that were sampled (out of 11 possible) and zz indicates whether the \mmb\ amplitude was evolving (ev) or not evolving (ne). All but two of the models use our fiducial GSMF from \citet{Leja_2020}. The remaining two of the models are based on an explicitly evolving form of the \citet{Liepold_2024} GSMF which we define in section \ref{sec:methods_gsmf}. }
\label{tab:models}
\end{deluxetable}

Because the models with evolving \mmb\ relations are a strict parametric super set of those without (i.e., the models are nested)\revii{, and since our prior for $\alpha_z$ is uninformative}, we performed Savage-Dickey ratio tests to quantify the significance of the evolving model over the fixed model \citep{Dickey_1971, Wagenmakers_2010}. For all our figures and calculations we present the results from fits to the first \fnum\ frequency bins. Our conclusions are not sensitive to the number of frequency bins we fit to.

\section{Results}\label{sec:results}

\begin{deluxetable}{lrrr}
\tablewidth{0pt}
\tablehead{
\colhead{Model Name} & \colhead{$\alpha_z$} & \colhead{\%  Positive} & \colhead{S-D Ratio}
}
\startdata
\asaev\ & 0.84 $\pm$ 0.79 & 87.2 & 1.3 \\
\asmev\ & 1.02 $\pm$ 0.48 & 99.3 & 24.6 \\
\asbev\ & 1.05 $\pm$ 0.55 & 96.8 & 5.2 \\
\liepev\ & 0.86 $\pm$ 0.76 & 88.7 & 1.5 \\
\enddata
\caption{The values of $\alpha_z$ and significance of each of our evolving models. Each quoted value for $\alpha_z$ is the median of the posterior and the associated error represents the 68\% confidence interval. We report the percentage of the posterior distributions that are positive alongside the associated Savage-Dickey (S-D) ratio between models with evolving and constant \mmb\ relations. The two models that fixed some / all GSMF parameters (\asmev\ / \asbev) show strong evidence for a positively evolving \mmb\ amplitude. Models that sampled all 11 GSMF parameters (\asaev\ and \liepev) did not converge to an equally constrained value for $\alpha_z$. These two models return lower values of $\alpha_z$ overall and are consistent with no significant evolution in the \mmb\ amplitude.}
\label{tab:alphaz}
\end{deluxetable}

\begin{figure}[ht]\centering
    \includegraphics[width=\columnwidth, keepaspectratio]{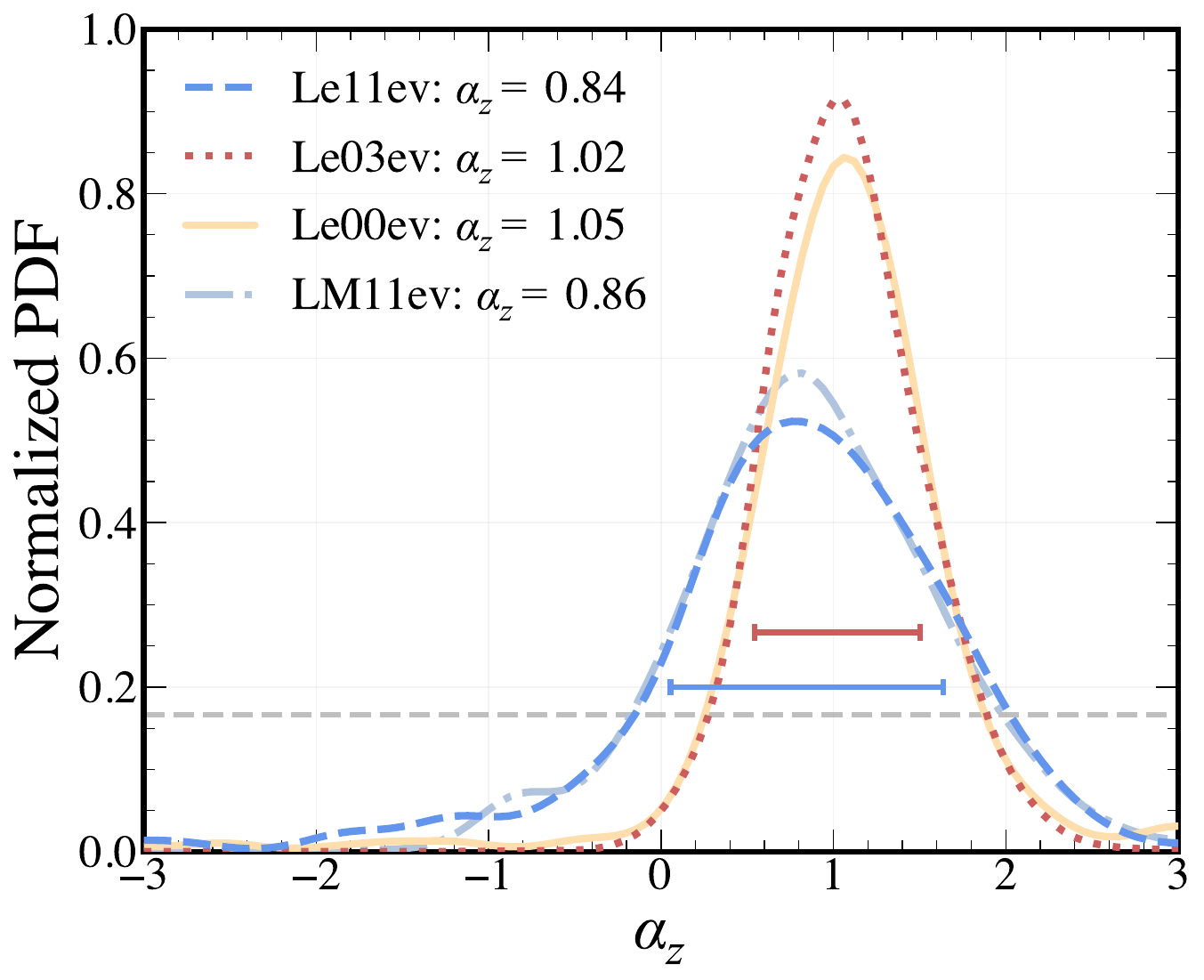}
    \caption{Here we show the four posterior distributions for $\alpha_z$ from our models. The \rev{upper} red and \rev{lower} blue horizontal lines indicate the 68\% confidence region for \asmev\ and \asaev. The gray dashed line shows our uniform prior. These are functionally identical to that of \liepev\ and \asbev\ respectively. In each distribution 87.2\%  - 99.3\% of values are positive, indicating a moderate to strong positive evolution in the \mmb\ amplitude. The posterior distributions for our models fall into two categories: (i) A wide range of $\alpha_z$ values with a significant (greater than 10\%) fraction of the distribution falling below  $\alpha_z = 0$ and (ii) A very narrow, nearly symmetrical, distribution of $\alpha_z$ values with only a negligible (under 5\%) fraction of the distribution sitting below $\alpha_z = 0$. This first category corresponds to models that sampled all 11 GSMF parameters (\asaev\ and \liepev), these distributions, while largely positive, are consistent with no significant redshift evolution of the \mmb\ relation. The latter category, however, show strong evidence for a positive \mmb\ amplitude evolution. The distributions from this second category had some or all of the GSMF parameters fixed (\asmev\ and \asbev) in the prior set up. With fewer degrees of freedom these models converged to higher values of $\alpha_z$ to a higher degree of confidence suggesting a better constraint for the \mmb\ amplitude evolution than the larger models. This is indicative of the degeneracy between the GSMF and \mmb\ parameters in our models.}
    \label{fig:alphaz}
\end{figure}

\begin{figure*}[ht]\centering
    \includegraphics[width=\textwidth, keepaspectratio]{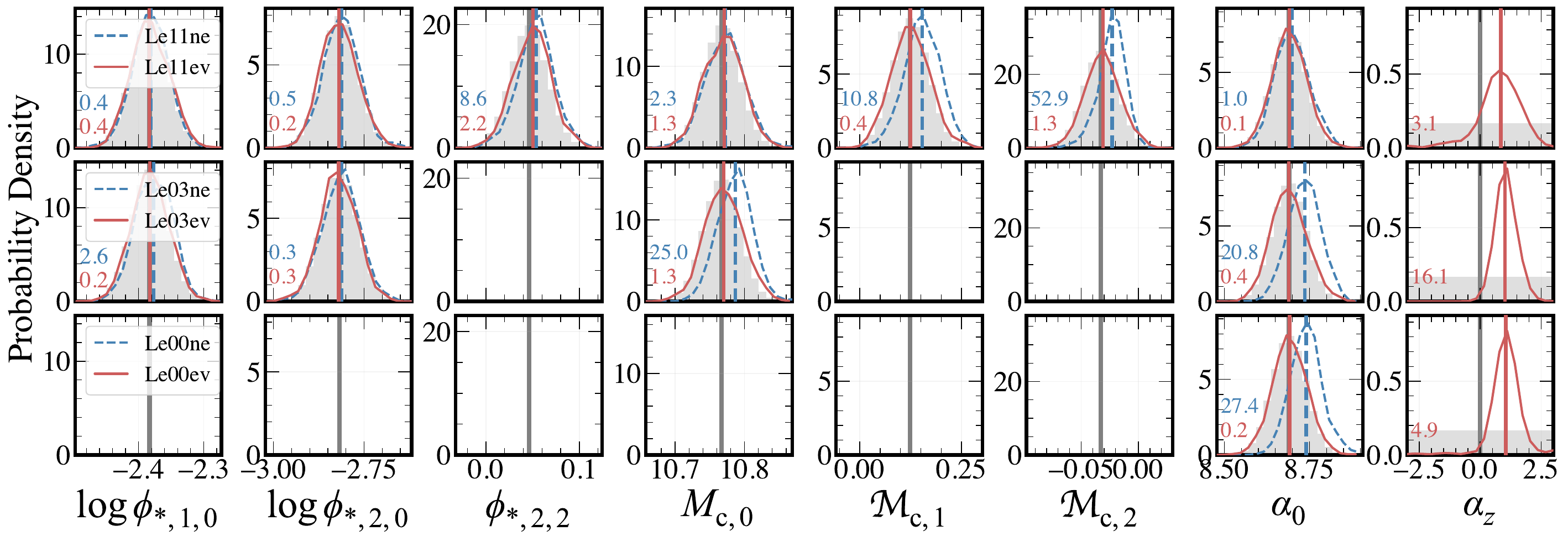}
    \caption{The posterior distributions for each of the evolving (red \rev{solid}) and non-evolving (blue \rev{dashed}) models. The priors are represented by the gray histograms in each panel and the gray vertical lines indicate the fiducial values. Each row represents a different evolving / non-evolving \mmb\ model pair---\textit{top:} \asa\ and \asaev, \textit{middle:} \asm\ and \asmev, \textit{bottom:} \asb\ and \asbev. In the case where a parameter was fixed, there is no histogram and only the vertical line is shown. \rev{In each panel we show the Kullback-Leibler divergence, $D_{KL}$, between each model posterior and the prior. The top, blue, number is for the Le\_ne models and the bottom, red, number is for Le\_ev models. For each parameter, $D_{KL}$ is equal to or lower for models which allow for \mmb\ evolution, indicating that the posterior distributions are in equal or better agreement with the priors when compared to the fixed $\alpha_z=0$ counterparts.} All three posterior distributions for $\alpha_z$ (right-most column) demonstrate a preference for positive values. The $\alpha_z$ posterior is relatively broad for the top row (\asaev), but when the evolutionary GSMF parameters are fixed (\asmev, middle row) the distribution shifts towards higher values and also narrows. When only the local GSMF parameters were sampled, but $\alpha_z$ was fixed to 0 (\asm), the posterior distributions deviated from the priors, not only for GSMF parameters (e.g, $M_{c, 0}$, $M_{c, 1}$, and $M_{c, 2}$), but also for $\alpha_0$. When $\alpha_z$ was allowed to vary, the posteriors recover the prior GSMF parameters. Additionally, when any number of the GSMF parameters are fixed (bottom two rows, \asmev\ and \asbev), the $\alpha_z =0$ models return high posterior values for the local \mmb\ amplitude, $\alpha_0$. This is indicative of the degeneracy between increasing galaxy number density and increasing the \mmb\ amplitude (either locally and/or at high-$z$). We also see little change between $\alpha_0$ and $\alpha_z$ posteriors in the middle and lower rows suggesting that fixing the three local GSMF parameters (\asbev) does not have a further affect over fixing only the evolutionary parameters. Overall, the models that allow the \mmb\ relation to evolve are otherwise in better agreement with observational constraints for the GSMF and local \mmb\ amplitude and have posterior distributions for $\alpha_z$ that are between 87 and 99\% positive.}
    \label{fig:large_posterior_comp}
\end{figure*}

A summary of the results for the posterior values for $\alpha_z$ are shown in Table \ref{tab:alphaz} and Figure \ref{fig:alphaz}. \rev{In Figure \ref{fig:large_posterior_comp} we highlight parameters of interest for which we calculate Kullback-Leibler divergences, $D_{KL}$, between posterior and prior distributions \citep{Kullback_1951}.} The best-fit spectra for all our models are presented in Figure \ref{fig:all_hcvsf_comp}. We find that, in general, a large number of massive SMBHs are needed to reproduce the GWB. When $\alpha_z$ is allowed to vary, the majority of the posterior distribution is positive with a median value of $\alpha_z \sim \resalphaall$ across all models and $\alpha_z \sim \resalpha$ for models with statistically significant evolution. We additionally find that, for models with $\alpha_z$ fixed to be 0, this high number density of massive SMBHs can be modeled either through a top-heavy GSMF (between 0.2 to 3 dex higher in number density for $M_\star \geq 10^{11} M_\odot$ and $z > 1$ compared to the observed GSMF, see Figures \ref{fig:gsmf_post_diff} and \ref{fig:leja_liepold}) or a high \mmb\ amplitude \citep[with $\alpha_0$ typically $\sim 8.89 \pm 0.11$ compared to the locally observed value of 8.69,][]{Kormendy_Ho_2013}. We first present our full-sample model containing posterior distributions and best-fit spectra for all 29 (30 with $\alpha_z$) parameters. Then, to highlight the effects of the degeneracy between parameters, we repeat this process with a subset of sampled parameters.

In section \ref{sec:res_gsmf_choice} we present our analysis of the GSMF parameters which we then further discuss in section \ref{sec:discussion}.

\subsection{Full-Sample Models}\label{sec:fullsamp}

Here we present the results of models that sampled all 29 (30 with $\alpha_z$) parameters. First we describe the outcome of the models that used our fiducial GSMF from \citet{Leja_2020}, then we present the impact of our test cases for explicitly evolving GSMF based on \citet{Liepold_2024}.

\subsubsection{\asa\ and \asaev}

For the majority of parameters, the posterior distributions recover the priors for both \asa\ and \asaev. In the top row of Figure \ref{fig:large_posterior_comp} we highlight eight parameters from these models; \rev{we discuss the effects of the remaining parameters in appendix \ref{sec:app_comp}} and complete corner plots can be found in Appendix \ref{sec:app_corner}. The non-evolving model, \asa\, shows slight deviations towards higher values for three of the the GSMF characteristic mass and normalization evolutionary parameters ($\phi_{*, 2, 2}$, $M_{c, 1}$, and $M_{c, 2}$). The two posterior distributions with the largest deviations from the priors are both of the evolutionary parameters for the characteristic stellar mass. When the \mmb\ relation is allowed to evolve, the posterior distributions for \asaev\ are more consistent with the priors for all GSMF parameters, though there is still a mild offset in the same direction as \asa\ for $\phi_{*, 2, 2}$. The posterior distribution for $\alpha_z$ is 87.2\% positive with a median value of 0.85. A Savage-Dickey ratio test in favor of the evolving model returns a value of 1.3, which indicates that this fit is consistent with no \mmb\ evolution.

The local GSMF values are recovered in both models, but the non-evolving model posteriors return generally larger values for the  $M_c$ evolutionary parameters. Higher values for $M_{c, 1}$ and $M_{c, 2}$ produce greater number densities of galaxies at higher redshifts, while maintaining consistency with the fiducial local number density. In particular, larger values for the characteristic mass produce more massive galaxies and therefore a greater number of massive SMBHs.

\rev{We use $D_{KL}$ to quantify the deviation of the posteriors from the priors for \asa\ and \asaev. Two distributions with $D_{KL} = 0$ are equivalent while $D_{KL} > 0$ indicate disagreement. For all the parameters shown in Figure \ref{fig:large_posterior_comp}, the value of $D_{KL}$ for \asa\ is equivalent or greater than for \asaev. This means that the posterior distributions for \asa\ diverge from the priors more than in \asaev\ suggesting that the posterior distributions for \asaev\ are in better agreement with the observed GSMF and local \mmb\ relation.}

Figure \ref{fig:gsmf_post_diff} (see also Fig.\ \ref{fig:leja_liepold}) compares the GSMFs from \citet{Leja_2020} to the GSMF implied by the median posterior values from \asa\ and \asaev. In the local universe, all three GSMFs are in good agreement. By redshift 2, both GSMFs from \asa\ and \asaev\ lie above the observed values, but \asaev\ is consistent, within error bars with \citet{Leja_2020}. The difference between the GSMFs becomes larger with redshift with \asaev\ remaining consistent with the observed GSMF while \asa\ is 0.5 to 2 orders of magnitude greater for galaxies with $ M_\star \geq 10^{10.5}\ M_\odot$ at $z = 3$. These massive galaxies are the hosts to the SMBHs in the mass range that dominates the GWB signal \citep[$M_{BH} > 10^9\ M_\odot$ ][]{AgazieBHB_2023}. When predicting SMBH masses using the local \mmb\ relation, to increase the amplitude of the modeled GWB, we need an increased number density of these massive galaxies compared to the fiducial model. Alternatively, an evolving \mmb\ relation offers a way to increase the number density of the most massive SMBHs without any changes to the galaxy population or the locally observed SMBH--galaxy relation. This suggests that the best-fit parameters for models with an evolving \mmb\ amplitude are more consistent with observational constraints than non-evolving models.

\begin{figure*}[ht]\centering
    \includegraphics[width=\textwidth, keepaspectratio]{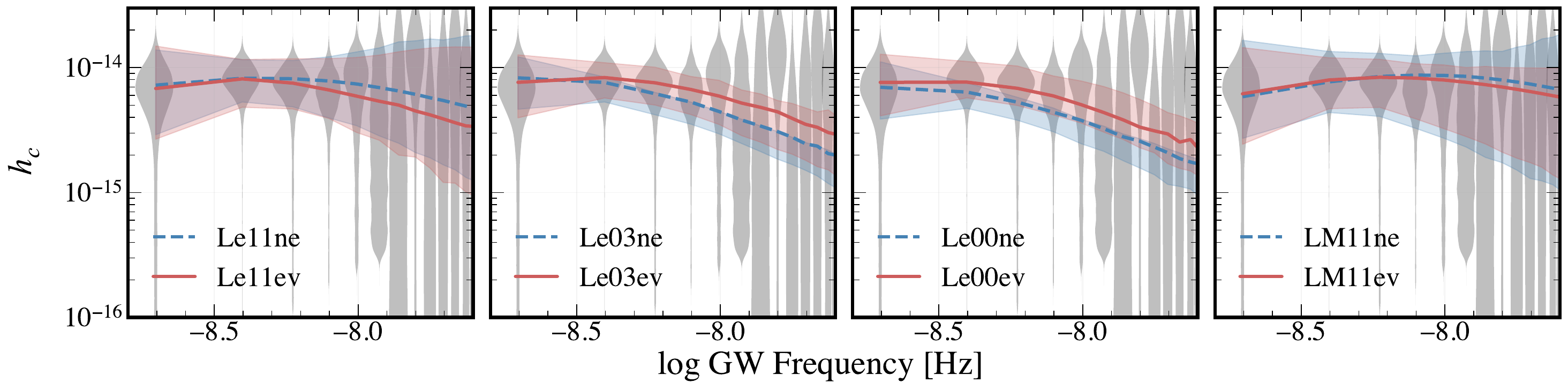}
    \caption{The GWB spectra associated with the best-fit parameters from our \modelnum\ models fit to the first five frequency bins. \rev{The blue dashed lines indicate models with $\alpha_z$ fixed to 0 and red solid lines indicate models which allowed for \mmb\ evolution.} In all four panels, the spectrum from the evolving \mmb\ models are in good agreement with the data though we note that the likelihood value for \asbev\ is the lowest of the \modelnum\ models (discussed further in section \ref{sec:res_gsmf_choice} \rev{and appendix \ref{sec:app_comp}}). There are some differences between the slope of the high-frequency end, but this portion of the spectrum is poorly constrained and it is not possible to distinguish between goodness of fit in this regime at this time. When all GSMF parameters are sampled (left- and right-hand panels, \asa\ and \liep), the spectra are nearly identical between the evolving and non-evolving models. The models with some/all GSMF parameters fixed (\asm\ and \asb) show more significant differences between the evolving and non-evolving models. For these models, those that allowed $\alpha_z$ to vary are consistently in better agreement with the data than the fixed $\alpha_z = 0$ counterparts indicating that these evolving models are a better description of the GWB than their non-evolving counterparts.}
    \label{fig:all_hcvsf_comp}
\end{figure*}

\begin{figure*}[ht]\centering
    \includegraphics[width=\textwidth, keepaspectratio]{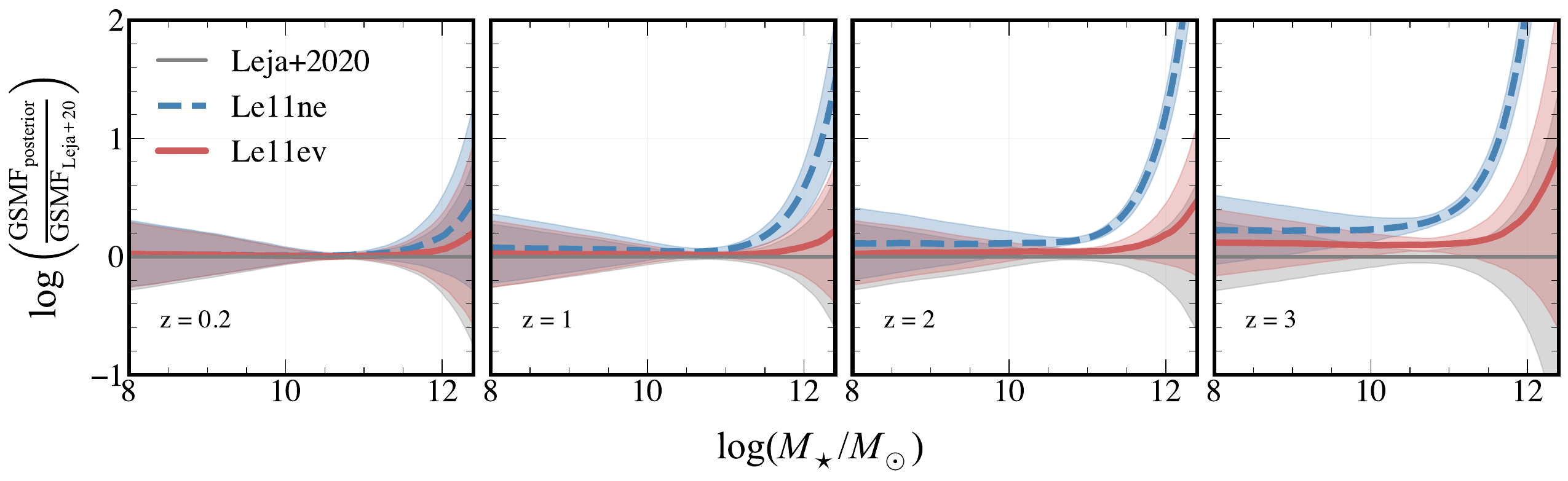}
    \caption{\rev{Here we plot the difference between the posterior GSMF from both the evolving (\asaev, red solid) and non-evolving (\asa, blue dashed) \mmb\ model and the observed GSMF from \citet{Leja_2020}. The light gray region represents the 1$\sigma$ error in the GSMF from \citet{Leja_2020}, we show the equivalent error range for \asa\ in blue and for \asaev\ in red. We compare the models only within the redshift range used in \citet{Leja_2020} $0.2 < z < 3$. We see that, when the \mmb\ relation is not allowed to evolve, the GSMF shows higher number densities of high-mass galaxies with an increasing discrepancy as redshift increases. This difference is highest for galaxies with $M_\star > 10^{11} M_\odot$ though by $z = 3$ the posterior GSMF from \asa\ is inconsistent with the observed GSMF at all masses $M_\star > 10^{9} M_\odot$. When the \mmb\ amplitude is allowed to evolve, however, the posterior GSMF is consistent within the uncertainties of the observed GSMF though with a minor positive offset. This difference is evidence that our best-fit models with an evolving \mmb\ amplitude are in better agreement with observational constraints for galaxy number density than the non-evolving models.}}
    \label{fig:gsmf_post_diff}
\end{figure*}

\subsubsection{\liep\ and \liepev} \label{sec:res_liep}

We find an anti-correlation between the strength of the GSMF evolution we assume and the resulting GWB amplitude. A strongly evolving GSMF (larger values of $M_{c, i}$) predicts lower mass densities of galaxies relative to a weakly evolving GSMF (lower values of $M_{c, i}$). The strongly evolving GSMFs also, therefore, produce lower number densities of the most massive SMBHs which then produces to lower GWB amplitudes.

Despite the greatly increased number density of local galaxies in this model, the strongly evolving GSMF produces a GWB spectrum that is only marginally greater in amplitude than the fiducial model. In fact, the posterior distributions, using the strongly evolving \citet{Liepold_2024} GSMF (with its much higher number density of galaxies with $M_\star \geq 10^{11.5} M_\odot$), display similar behavior to that of \asa\ and \asaev. That is, the GSMF posteriors for models without \mmb\ evolution tend towards greater number densities of high-mass galaxies for $z \gtrsim 1$. The GSMF posteriors recover the priors when \mmb\ evolution is allowed and the posterior distribution for $\alpha_z$ is nearly identical to that of \asaev\ (see Figure \ref{fig:alphaz}). 

In Figure \ref{fig:leja_liepold} we compare the prior and posterior GSMFs for all full-sample models. We see that the galaxy number densities for $1 \lesssim z < 3$ are roughly equivalent for \asa\ and \liep\ (similarly for \asaev\ and \liepev). We see a more significant difference between models that have an evolving or static \mmb\ relation than between models with different prior GSMFs (further discussed in section \ref{sec:res_gsmf_choice}). This result suggests that, to reproduce the GWB amplitude, we need a significantly greater number density of galaxies than we currently observe, not only locally, but at least out to $z = 3$. We discuss the physical implications and feasibility of this model further in section \ref{sec:discussion}.

While the local GSMF measured by \citet{Liepold_2024} offers valuable insight into the local galaxy population and the limits of volume-limited surveys, we do not find that this local increase in galaxy number density is sufficient to reproduce the observed GWB spectrum. The difference between our result and the what \citet{Liepold_2024} find is a direct result of the assumptions made about binary hardening mechanisms in our respective calculations. For their estimate of the GWB amplitude, \citet{Liepold_2024} follow the methods of \citet{Sato-Polito_2024} and \citet{Phinney_2001}. This model includes only the effect of gravitational wave-driven SMBH binary hardening, which produces a power-law spectrum. In this work, we assume that binaries can also harden through dynamical friction and stellar scattering. The effect of these additional hardening pathways is to change the shape of the overall spectrum and to lower the amplitude. While keeping all other parameters constant, the change in shape and amplitude has the greatest impact at the low-frequency end of the spectrum \citep[$f \lesssim 10 \mathrm{nHz}$, e.g., see lower-right panel of Figure 4 in][]{Agazie_2023}. The shape of the spectrum, when including these additional hardening mechanisms, provides better fits to the data across the observed range of frequencies than the gravitational wave-only model \citep{Agazie_2023}.

\subsection{Fixed GSMF Evolution Models: \asm\ and \asmev}\label{sec:redsampsm}

The first of the sub-models that we discuss are \asm\ and \asmev. Each of these is identical to \asa\ and \asaev\ respectively except the six evolutionary parameters ($\phi_{*, 1,1}$,  $\phi_{*, 1,2}$, $\phi_{*, 2,1}$,  $\phi_{*, 2,2}$, $M_{\mathrm{c} ,1}$, and $M_{\mathrm{c} ,2}$) alongside $\alpha_1$ and $\alpha_2$ for the GSMF are fixed to the posterior values given by \citet{Leja_2020}. To focus on the degeneracy between GSMF and \mmb\ parameters we additionally fixed all parameters for galaxy merger rates and bulge fractions to their fiducial values. These models, therefore, sample a total of eight parameters (nine with $\alpha_z$).

When sampling from all 11 GSMF parameters, the posteriors for the model that did not allow for \mmb\ evolution (\asaev) recovered the priors for the local GSMF values while only the evolutionary parameters deviated. Now, since these evolutionary parameters were fixed, the local GSMF posterior for characteristic mass is distributed towards greater values. Unlike before, this model also returns a posterior distribution with generally larger values of the \mmb\ amplitude than the prior. An increase to either the characteristic galaxy mass or the \mmb\ amplitude can generate the top-heavy BHMFs necessary to reproduce the observed GWB. A moderate (within 1$\sigma$) increase to both parameters, on the other hand, allows for massive SMBH populations to be modeled without deviating too far from either one parameter. This result is consistent with what \citet{AgazieBHB_2023} found---to match the GWB either a large number of parameters in the fiducial model need to change by a small amount, or a very small number of parameters must be significantly different.

When the \mmb\ relation is allowed to evolve, the posterior distributions for all GSMF parameters are in good agreement with the priors. Interestingly, the distribution for $\alpha_z$, for this model, is 99.3\% positive with a Savage-Dickey ratio of 24.6. The standard deviation of this distribution is also significantly reduced indicating a higher confidence of positive evolution.

\rev{Similarly to \asa\ versus \asaev, the $D_{KL}$ between the priors and posteriors for \asm\ is equivalent or greater than for \asmev. As mentioned earlier, the posterior distributions for $M_{\mathrm{c} ,0}$ and $\alpha_0$ diverge from the priors more significantly in \asm\ than in \asa. In \asa\ the two distributions with the highest divergence were $M_{\mathrm{c} ,1}$, and $M_{\mathrm{c} ,2}$. These values are fixed in \revii{\asm} and so this model compensates by increasing $M_{\mathrm{c} ,0}$, and $\alpha_0$. We do not see a notable change to the $D_{KL}$ values for these distributions in \asbev\ versus \asaev.}

Nearly all samples from the model that did not include \mmb\ evolution tend towards lower GWB amplitudes. \rev{In the second panel of Figure \ref{fig:all_hcvsf_comp} we see that the shape of the GWB spectrum is nearly straight. Except for the two right-most frequency bins, the best-fit spectrum is equivalent to the upper limit on the models meaning that no models produced a GWB amplitude higher than the best-fit model for the higher-frequencies. The best-fit model here is roughly consistent (though on the low side) with four of the five frequency bins we used for fitting.} The \asmev\ model, on the other hand, while lower in GWB amplitude than \asaev\ at the high frequency end, is still consistent with the PTA data.

\subsection{Fixed GSMF Models: \asb\ and \asbev}\label{sec:redsampf}

Finally, we present the smallest subset of our models, which only sample the three hardening parameters, the \mmb\ amplitude, and $\alpha_z$ (for \asbev). These models fix all 11 GSMF parameters to their fiducial values from \citet{Leja_2020} so the local and high-$z$ GSMF cannot vary for these models.

This subset of models returns a posterior $\alpha_z$ distribution that is 96.8\% positive with a Savage-Dickey ratio of 5.2. In this case, the posterior distribution for \mmb\ amplitude recovers the prior almost exactly. For the non-evolving model, to reproduce high SMBH masses, a greater overall \mmb\ amplitude is required. This result demonstrates how the models will trend more strongly towards positive values of $\alpha_z$ as the other options for increasing the number density of massive SMBHs are removed. When the option for evolution is also removed, the posteriors will deviate strongly from the priors for parameters that are influential on the predicted SMBH number density.

The GWB spectra share similar characteristics to the \asm\ and \asmev\ models with the non-\mmb\ evolving models consistently lying below the majority of the PTA data and the evolving \mmb\ models falling closer to the center of the violins in more bins, but lying slightly above the median. We note that the best-fit spectrum for \rev{\asb\ }has the lowest likelihood of all \modelnum\ spectra \revii{indicating it is the worst fit to the GWB}, an effect we detail further in Appendix \ref{sec:app_comp}.

\subsection{Influence of the GSMF} \label{sec:res_gsmf_choice}

\begin{figure*}[ht]\centering
    \includegraphics[width=\textwidth, keepaspectratio]{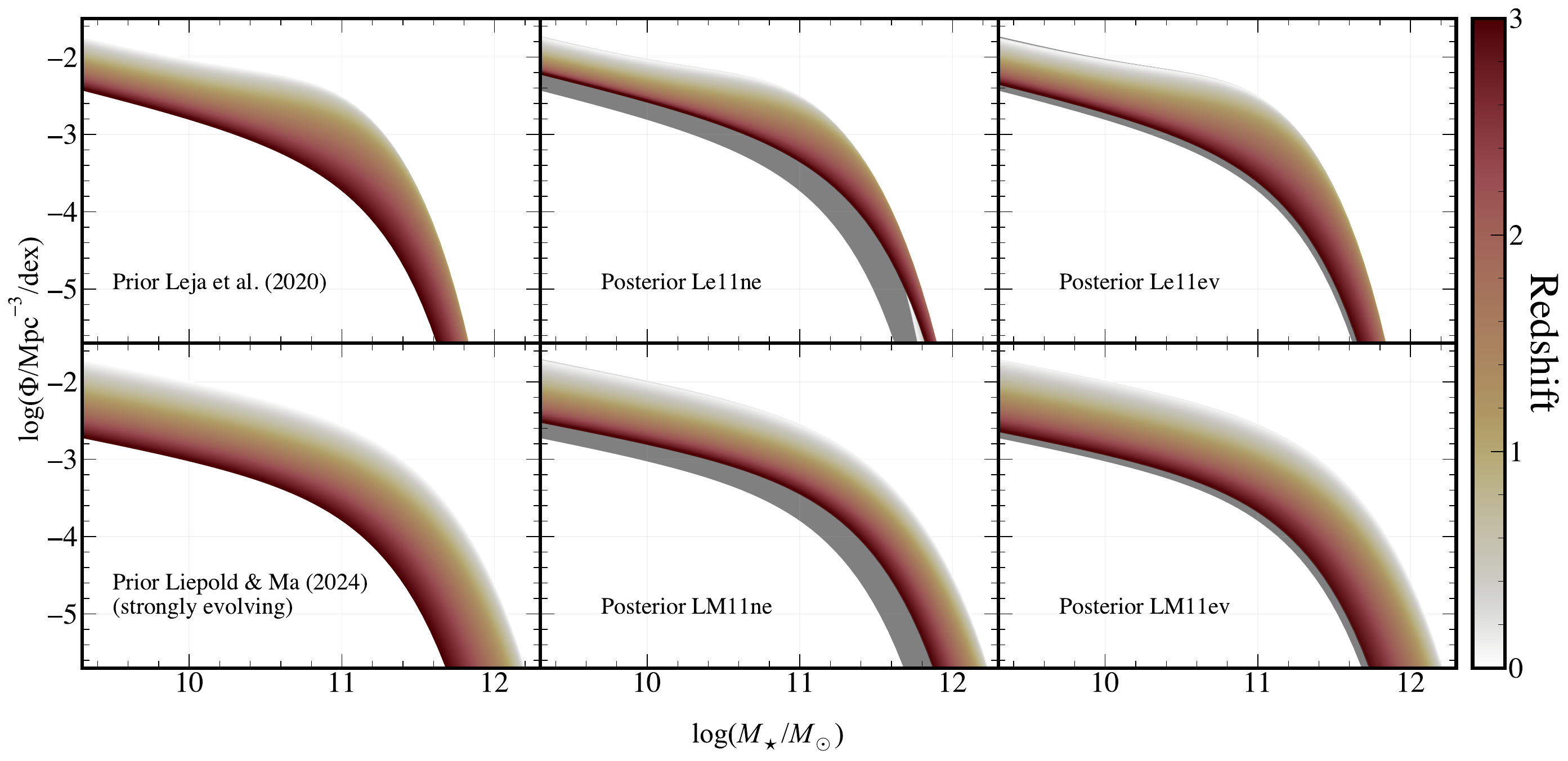}
    \caption{\textit{Top left:} The prior GSMF from \citet{Leja_2020}. \textit{Top middle:} The posterior GSMF for \asa. \rev{\textit{Top right:}} The posterior GSMF for \asaev.
    \textit{Bottom:} Same as top row, but for models based on our \rev{strongly} evolving version of the \citet{Liepold_2024} GSMF. We see that, in both middle plots, the models that have $\alpha_z = 0$ require a large number density of galaxies relative to their respective fiducial model. Despite the increased local number density, the posterior GSMF from \liep\ returns a greater number density for $z = 3$ compared to the \citet{Leja_2020} GSMF. The prior for this model had a similar number density to that of \citet{Leja_2020} in this redshift range, by design. The number densities for $2 \lesssim z \lesssim 3$ are similar between the two posterior models with $\alpha_z = 0$. This means that, to reproduce the GWB without an evolving \mmb\ amplitude, we need a significantly higher number density of massive galaxies across out to $z \sim 3$. For both models that allowed $\alpha_z$ to vary (right-most panels) the posterior GSMFs are only negligibly different from the priors and the corresponding median posterior values for $\alpha_z$ are similar (see Figure \ref{fig:alphaz} and Table \ref{tab:alphaz}) implying that the locally increased number density for \liepev\ does not sufficiently increase the GWB amplitude without additionally increasing the high-$z$ GSMF number density or the \mmb\ amplitude.}
    \label{fig:leja_liepold}
\end{figure*}

\begin{figure}[ht]\centering
    \includegraphics[width=\columnwidth, keepaspectratio]{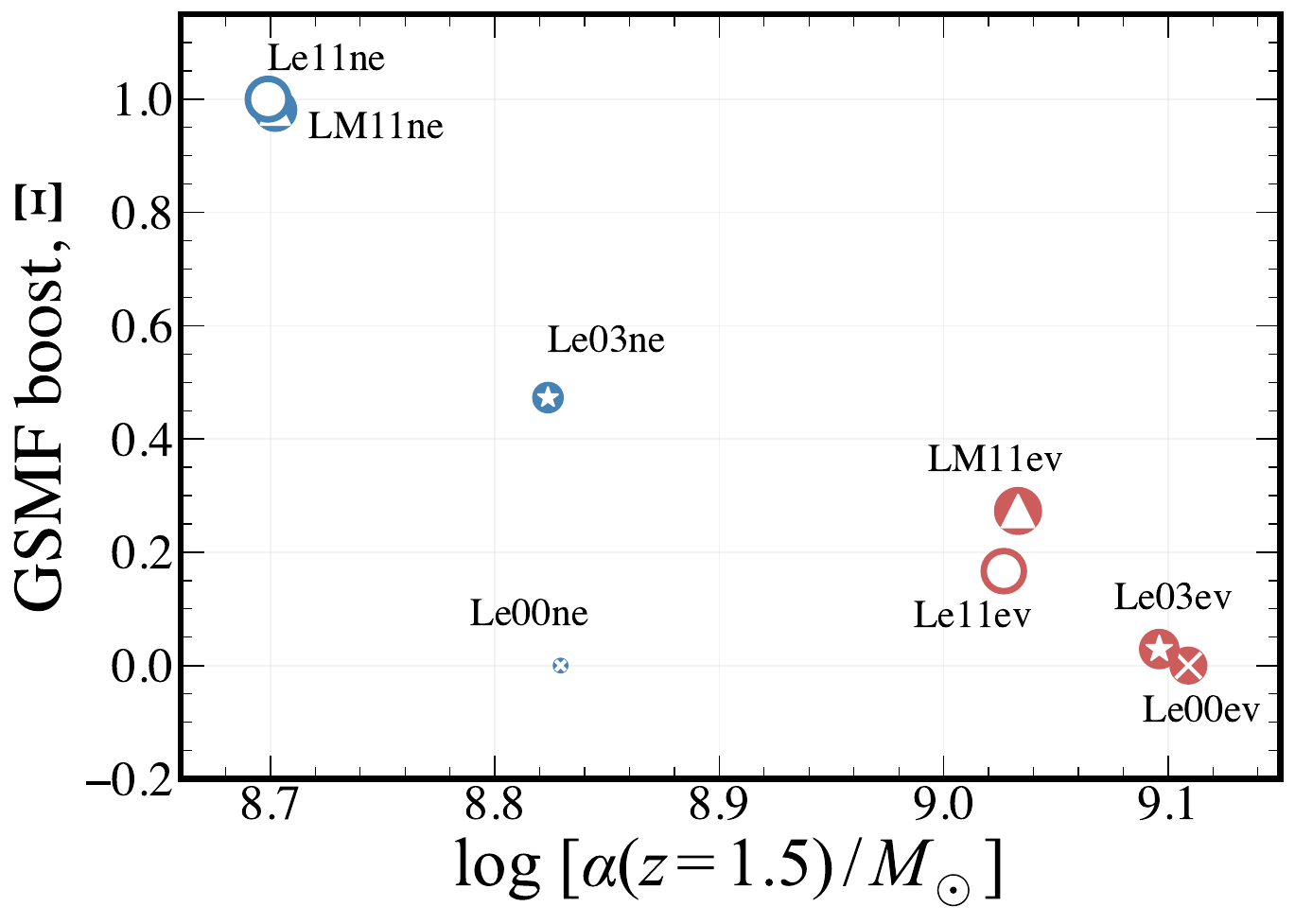}
    \caption{The quantity, $\Xi$, represents the normalized number density ``boost'' in a given posterior GSMF relative to the corresponding fiducial GSMF for $11 < \log M_\star < 13$ and $0.5 < z < 3$. $\Xi = 0$ means the two GSMFs are equivalent, a positive value of $\Xi$ corresponds to an increased number density relative to the fiducial model across our integration range, all $\Xi$ values were normalized to a maximum of 1. Red points are for models that allow the \mmb\ relation to evolve and blue points fixed $\alpha_z$ to be 0. \revii{The size of the points corresponds to the log likelihood value of the best-fit GWB spectrum. Larger circles indicate better fits to the GWB data while low values of $\Xi$ indicate better agreement with the GSMF. Therefore models which are consistent with both datasets have large marker sizes and are low on the y-axis.} Each model pair shares a marker style overlaid in white (e.g., \asa\ and \asaev\ are blue and red circles with stars). There is a clear negative correlation between pairs and also across all models. This demonstrates the degeneracy between an increased number density of galaxies and an increased SMBH--galaxy mass ratio. This degeneracy exists because both are valid pathways to producing the high number density of massive SMBHs implied by the GWB.}
    \label{fig:xi_vs_alpha}
\end{figure}

In this section we present the posterior GSMFs in two ways. We show the posterior GSMFs over-plotted with the corresponding prior GSMF for full GSMF sample models in Figure \ref{fig:leja_liepold} for visual comparison. We then quantify the number density increase relative to \citet{Leja_2020}, which we plot versus the corresponding $\alpha(z)$ posterior median to demonstrate the degeneracy between these quantities in Figure \ref{fig:xi_vs_alpha}.

In Figure \ref{fig:leja_liepold} we can see that the posterior GSMF from \asa\ and \liep\ generally have higher number densities than the fiducial model. This offset is most extreme for the \liep\ GSMF which is to be expected since the fiducial GSMF for this model started with a higher number density than the \citet{Leja_2020} GSMF for $z < 1$. Both of these GSMFs are greater in number density than their respective fiducial model and the fixed $\alpha_z = 0$ counterpart. Though we highlight the most extreme examples here, this is a trend across models that the fixed $\alpha_z = 0$ models require a greater number density of (especially the most massive) galaxies.

Because the GSMF is defined by up to 11 parameters in our models, the degeneracy between $\alpha(z)$ and any given GSMF parameter is not very strong. There is, however, considerable degeneracy between the \mmb\ evolution and an increased number density of galaxies. To further demonstrate this degeneracy we define the quantity $\Xi$ to be the integrated difference of the number density of galaxies between two models (see equation \ref{eqn:xi}). We integrate over $11 < \log M_\star < 13$ in mass and $0.5 \leq z \leq 3$ in redshift and note that trend of the results of this analysis are not sensitive to the bounds of our integral. We chose to omit $0 < z < 0.5$ in the integration range for clarity, including this range only results in a constant positive offset in $\Xi$ for both \liep\ and \liepev\. We additionally normalize $\Xi$ such that all our values lie between 0 and 1; $\Xi  = 0$ indicates that the posterior GSMF is equivalent to the prior GSMF across our integration range (which is the case for both \asb\ and \asbev). The $\Xi$ parameter encapsulates the ``boost'' in galaxy number density and is analogous to the way $\alpha(z)$ encodes the changing \mmb\ amplitude.

\begin{equation}
    \Xi = \iint \left[ \phi_\mathrm{posterior}(M, z) - \phi_\mathrm{Leja+20}(M, z) \right] M dM dz
	\label{eqn:xi}
\end{equation}

In Figure \ref{fig:xi_vs_alpha} we plot $\Xi$ against the \mmb\ amplitude at a fixed redshift $\alpha(z = 1.5)$, locally all values of $\alpha(z)$ are equivalent and the value of $z$ we choose only shifts the evolving \mmb\ models left or right on the x-axis. The color differentiates evolving (red) models from non-evolving (blue) and the marker styles show the pairs of models with the same prior set up aside from \mmb\ evolution (i.e., each marker style appears twice, one on red and one on blue). The clear negative trend demonstrates the degeneracy between an increased number density of galaxies and an increased SMBH--galaxy mass ratio in our posterior models. In the cases where the GSMF was allowed to vary, the evolving models (\asaev,  \asmev, \asbev, and \liepev) have a lower value for $\Xi$ and a higher value for $\alpha(z=1.5)$ than their non-evolving counterparts (\asa\ and \asm). These models follow the same trend, but are additionally boosted relative to the fiducial \citet{Leja_2020} GSMF by design (as described in section \ref{sec:methods_gsmf} and Appendix \ref{sec:app_gsmf}). The only outlier to this trend is \asb. The best-fit likelihood value for this model is several orders of magnitude smaller than for other models (represented by the size of the points)\revii{ indicating it is a poor fit to the GWB data. While in good agreement with the GSMF (low $\Xi$) its location on the plot is not an accurate reflection of the overall trend since the trend is defined by models that can reproduce the GWB.} Therefore we conclude that the high number density of massive SMBHs, required to reproduce the GWB, can be produced either by a large increase to galaxy number density (GSMF), a large increase to the SMBH--galaxy mass ratio, or a moderate increase to both.

The models that fix some / all GSMF parameters converge to higher values of $\alpha_z$ with a greater degree of confidence. The \asmev\ model only fixed the evolutionary parameters from \citet{Leja_2020} and sampled the local GSMF values. The posteriors for this model recovered the local GSMF and \mmb\ parameters while its counterpart model, with $\alpha_z$ fixed to 0, (\asm) did not. This suggests that the GWB free-spectrum is well described by a SMBH population as predicted from an evolving \mmb\ relation and the galaxy number density given by the GSMF due to \citet{Leja_2020}. The GWB spectrum is equally well described by an unchanging \mmb\ relation if, and only if, the GSMF is significantly higher in number density (for $M_\star > 10^{11.5} M_\odot$), not only locally, but also out to $z$ as high as 3, which is not well supported by observational data \citep[e.g.,][]{Muzzin_2013, Tomczak_2014, Davidzon_2017, Behroozi_2020}, though it is not entirely ruled out \citep[e.g.,][]{Moustakas_2013, Wright_2018}.

\subsection{Black Hole Number Density}

\begin{figure*}[ht]\centering
    \includegraphics[width=\textwidth, keepaspectratio]{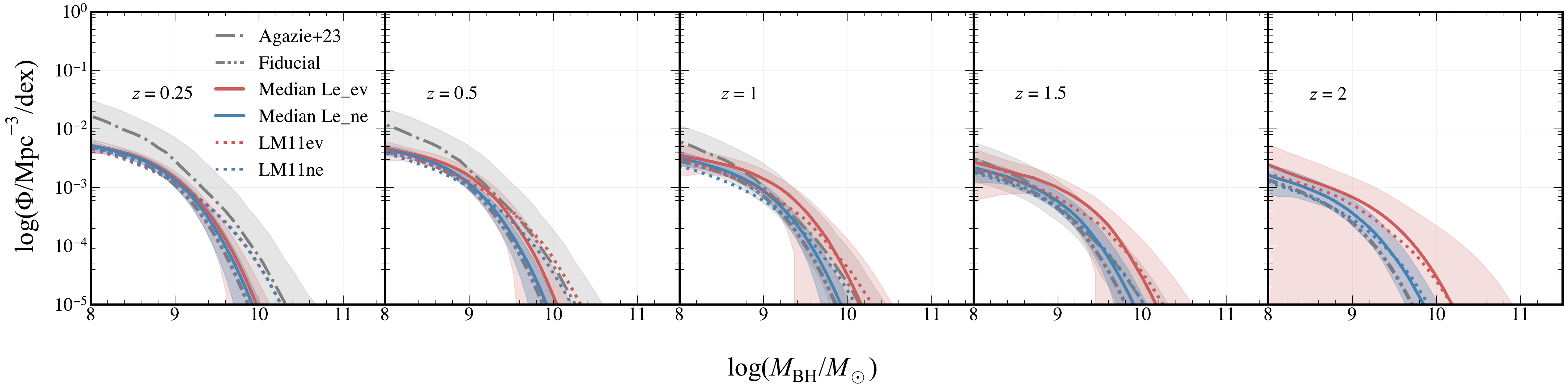}
    \caption{Posterior BHMFs from our best-fit models compared with the fiducial model and the BHMF from \citet{AgazieBHB_2023}. The red (blue) solid lines represent the median BHMF from all the evolving (non-evolving) models that used the \citet{Leja_2020} GSMF. Models using the \citet{Liepold_2024} follow this same color convention and are dotted instead of solid. We exclude \asb\ from these calculations because of its low likelihood value. Generally, the models that have an evolving \mmb\ have higher number densities at higher redshifts compared to the non-evolving counterparts. The models that used the \citet{Leja_2020} and did not allow for \mmb\ evolution are mildly positively offset from the fiducial BHMF. This offset increases with redshift reflecting the difference in posterior GSMF evolution.}
    \label{fig:bhmf}
\end{figure*}

In Figure \ref{fig:bhmf} we present the BHMFs resulting from our models for $0.25 \leq z \leq 2$. We additionally plot the posterior single SMBH functions from Figure 13 in \citet{AgazieBHB_2023} and our fiducial BHMF for comparison. The BHMFs for evolving models using the \citet{Leja_2020} GSMF (\asaev, \asmev, and \asbev) are nearly identical to each other and so we plot one BHMF that is the median of all three models. We compare this to the posterior BHMF from \liepev\ and find that the BHMFs are broadly consistent for $z > 0.5$, but that of \liepev\ is greater in number density for the most massive SMBHs for $z < 0.5$. This is to be expected since this is the same behavior seen in the respective GSMFs.

We also show the median BHMF for the fixed $\alpha_z = 0$ models, \asa\ and \asm, (excluding \asb\ due to its low likelihood). This BHMF has a lower number density than that of the evolving models across the entire mass range for $z > 1$. This BHMF is more similar to the fiducial BHMF for low redshifts, but has a boosted number density which increases with redshift. This increasing positive offset is reflective of the GSMF posteriors and the increased number density of galaxies in the non-evolving best-fit models (for \asa\ and \asm).

Generally, all posterior models have a higher number density of SMBHs relative to the fiducial model across all redshift bins while evolving models typically have the highest number densities in a given bin.

\section{Discussion}\label{sec:discussion}

Our results suggest that either the $z = 0$ \mmb\ relation does not apply at higher redshift or galaxy observational surveys drastically underestimate the number of massive galaxies at all redshifts $z < 3$ either through non-detection of a population or by underestimating masses. The best-fit evolving models indicate a positive offset for the \mmb\ amplitude in the past. An \mmb\ amplitude that increases with redshift ($\alpha_z > 0$) implies a black-hole-first evolution. Some studies have proposed this as a possible formation route stating that kinetic-mode feedback from SMBH accretion disks can impede the star formation within the host galaxy thus leading to a delayed stellar mass increase \citep{Zhuang_2023}, also known as black hole ``dominance'' \citep{Volonteri_2012}. In the opposite case, decreasing amplitude at high-$z$ may be indicative of radiative-mode feedback, where UV radiation from star formation prevents gas from sinking to the center of the galaxy therefore postponing SMBH accretion \citep{Zhuang_2023}. Our results support the black hole dominance model of SMBH--galaxy coevolution. The value for the \mmb\ amplitude evolution we find $\alpha_z = \resalpha \pm \restd$ is consistent with several observational \citep{Ding_2020, Pacucci_2023, Zhang_2023, Tanaka_2025} studies and some theoretical / simulation studies \citep[e.g., ][]{Wyithe_2003, Cattaneo_2005, Croton_2006}, but there is a lack of consensus among these studies, especially for the redshifts most relevant for the GWB \citep[e.g., ][]{Merloni_2010, Bennert_2011, Li_2021, Cisternas_2011}. Our analysis places the strongest constraints only on the most massive SMBHs ($M_\mathrm{BH} \geq 10^{8} M_\odot$). We cannot firmly differentiate between our power-law evolution and more complex models \citep[such as dual sequence behavior][]{Shimizu_2024} nor can we determine whether the evolution is mass-dependent \citep[e.g.,][]{Hoshi_2024}. A complementary follow-up analysis could consider the effect of heavy versus light SMBH seeding models on the low-mass end of the \mmb\ relation to form a bigger picture of \mmb\ evolution.

\begin{figure*}[ht]\centering
    \includegraphics[width=\textwidth, keepaspectratio]{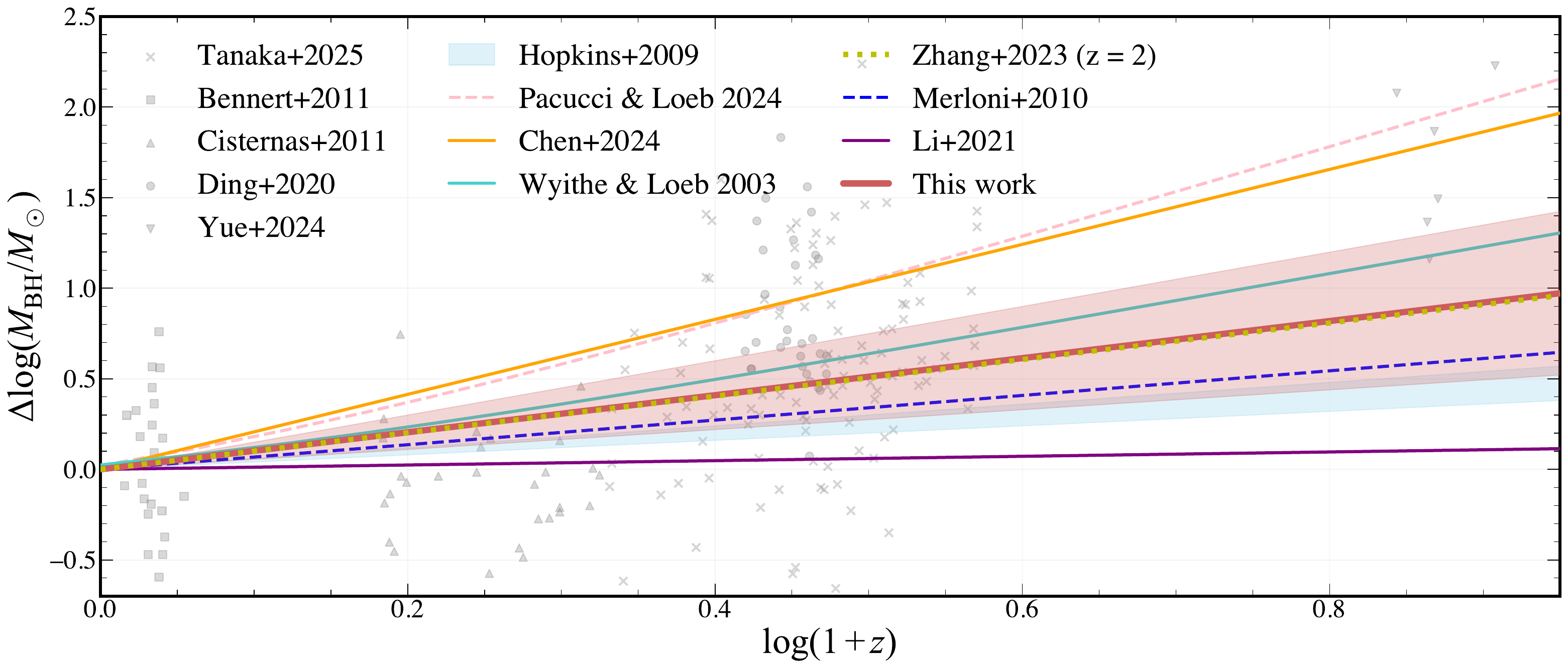}
    \caption{A comparison between different literature estimates of the \mmb\ relation and our results (thick, solid red line and 68\% confidence shaded region). The y-axis is the (log) offset between SMBH mass at a given redshift compared to the local amplitude. For this figure, we only include our models with statistically significant evolution (\asmev\ and \asbev). The gray points represent data while the different lines are fitting results. For \citet{Zhang_2023} (yellow dotted) we approximate the evolutionary form based on their reported redshift range and \mmb\ offset. Our analysis favors an intermediate positive evolution which is nearly identical to that of \citet{Zhang_2023} and similar to those of \citet{Wyithe_2003} and \citet{Merloni_2010}. Our model does not predict offsets as great as \citet{Pacucci_2024}, \citet{Yue_2024}, or \citet{Chen_2024}.}
    \label{fig:Mbh_Mstar}
\end{figure*}

In Figure \ref{fig:Mbh_Mstar}, we show the \mmb\ amplitude offset ($\Delta \log M_\mathrm{BH}/M_\odot$) versus redshift predicted by our model alongside measurements from literature. The gray scattered points represent individual galaxies with both stellar and SMBH mass measurements and so we calculate the offset from the local \mmb\ relation for these objects. The lines each represent a model from papers which provide an explicit functional form or offset for \mmb\ evolution. The thicker red line and shaded red region represent the median posterior value for $\alpha_z$ and 68\% confidence region from our models with statistically significant evolution (\asmev\ and \asbev). We find that our model is most consistent with that from \citet{Zhang_2023} (yellow dotted line) though our error bars encompass the models due to \citet{Wyithe_2003} and \citet{Merloni_2010}. We note that both \citet{Zhuang_2023} and \citet{Merloni_2010} were careful to account for observational bias \citep[e.g., as described by][]{Lauer_2007} in their analysis. Our model does not predict as extreme an offset as measured by \citet{Ding_2020}, \citet{Pacucci_2024}, or \citet{Yue_2024}.

We find $\alpha_z = \resalpha \pm \restd$ which lies below the value ($\alpha_z \sim 2.07 \pm 0.47$) found by \citet{Chen_2024}. \citet{Chen_2024} perform a similar analysis to the one we present here in which they fit the same data and also use the same fiducial \mmb\ relation as us. There are two key differences in their model versus ours: (i) the GSMF and (ii) the SMBH binary hardening model. They use the GSMF due to \citet{Behroozi_2020}, which has notably different (both higher and lower) number densities to our fiducial model across the mass and redshift ranges we consider. Because the GSMF they assume is not uniformly offset from ours, the effect on their $\alpha_z$ estimate is not obvious, but this difference is certainly a contributing factor to this discrepancy. Another difference between our models is the hardening prescription for the SMBH binaries. The model they use is a more complex model than we used here \citep{CYL_2020}. Differences in binary hardening efficiency can affect the GWB amplitude. We are able to model the GWB using $\alpha_z \sim 2.07$ if we assume a hardening timescale that is much longer than currently supported values \citep[by up to a factor of ten times higher than reported in][]{AgazieBHB_2023}. For this work, we assume a constant hardening timescale and the discrepancy between these results is demonstrative of the need for realistic binary hardening models. Future testing with EM studies will be important for placing bounds on hardening timescales. Inferences for SMBH populations from the GWB are sensitive to the underlying models used for galaxy populations. The difference in our results demonstrates the importance for having robust and well-constrained models.

Our analysis here suggests that, if we can reproduce the high GWB amplitude through only changes to the GSMF, then this would require a significantly higher number density of the highest mass galaxies at higher redshifts (at least out to $z = 3$). As they discuss in their paper, the increased number density that \citet{Liepold_2024} find indicates that galaxy surveys are missing out on the population of the most massive galaxies ($M_\star \geq 10^{11.5} M_\odot$) and/or the mass to light ratios underestimate galaxy mass for this population. Local surveys may miss out on this population of galaxies because the survey area is simply not large enough to include this otherwise rare population. For higher redshifts, however, the survey volume is sufficient to ensure completeness and so the, e.g., COSMOS \citep{Laigle_2016} and 3D-HST+CANDELS survey \citep{Skelton_2014} surveys used in \citet{Leja_2020} should be representative of the underlying population. This, combined with the abundance of reliable data, would suggest that the posterior GSMF at $z > 1$, needed by our models to match the GWB spectrum, is inconsistent with the true galaxy population. If, however, the initial mass function is bottom heavy, like that assumed by \citet{Liepold_2024} and minimally evolving (and therefore bottom heavy at higher redshifts), then a GSMF like that in the bottom middle panel of Figure \ref{fig:leja_liepold} could be feasible. In either case, analyses of the GWB are most directly a probe of SMBH properties, so any inferences we draw about the GSMF are several steps removed and are therefore only implicit. It is still useful to discuss the degeneracies with the GSMF in our models, however the GSMF estimates from this study should be treated with caution.

Gravitational wave-based studies, such as this, offer a new probe of SMBH--galaxy- mass scaling relations that can complement observational studies. Broadly speaking, there are two ways to observationally infer the SMBH mass function over cosmic time using only electromagnetic data: (i) BH mass scaling relations combined with observations of galaxies and (ii) models of Eddington ratio distributions combined with observations of AGN luminosity functions. These two methods have inferred incompatible SMBH mass functions, with galaxy-observation methods predicting higher densities than AGN-observation methods. A series of works \citep{Shankar_2016, Barausse_2017, Shankar_2019} proposed a potential solution whereby the (non-evolving) mass scaling relations had much lower amplitude (possibly a result of unsubstantiated measurement biases) to lower the galaxy-observation results to be in line with the AGN-observation results. Our work here shows that the GWB spectrum is generally in line with the galaxy-observation results. Not only are the measured scaling relations not too low at $z = 0$, they are, if anything, higher amplitude at higher redshift, something which is also noted by other studies \citep[e.g., ][]{Sato-Polito_2024, Liepold_2024}. One as yet under-explored possibility to reconcile these two observational results is that radiative efficiencies are lower by a factor of 2 -- 5 than typically assumed for the brightest AGN. This is outside the scope of this paper, but will be investigated in future studies.

Our analysis of the \mmb\ relation focused on evolution of the amplitude, however it is possible that the relation may be evolving in other ways. Recent work has suggested that a variety of growth pathways are likely to be relevant within galaxy populations and that they are encoded in the intrinsic scatter \citep{Zhuang_2023, Terrazas_2024, Hu_2025, Cohn_2025} where the local \mmb\ relation acts as a sort of ``attractor'' for SMBH--galaxy pairs as they evolve. This concept is not new \citep[e.g.,][]{Peng_2007, Hirschmann_2010, Merloni_2010, Jahnke_2011}, but has seen a lot of new analysis, especially in the last two years. In particular, \citet{Zhuang_2023} conducted a study of $z \leq 0.35$ AGNs and found that the scatter in the \mmb\ relation is greater in the early universe, a finding substantiated recently by both gravitational wave and simulation analyses \citep[e.g.,][]{Gardiner_2023, Terrazas_2024}. Furthermore, \citet{Li_2025} measured the SMBH--galaxy mass ratios for AGN at $z \sim 3-5$ and found that they are consistent with the local population. Results such as theirs could be indicative of an increased scatter in the \mmb\ relation outside the local universe. Biased observations of a high-scatter \mmb\ relation could appear, artificially, as an inflated amplitude. Future GWB studies similar to this will provide valuable insights into alternative evolutionary forms of the \mmb\ relation.

It is also possible that a scaling relation not based on galaxy mass may more accurately reproduce the highest-mass SMBH number density. It has been shown that the \mmb\ and \msigma\ relations predict different BHMFs outside the local universe \citep{Matt_2023} and that this has an impact on the predicted GWB amplitude \citep{Simon_2023}. On the other hand, \citet{McConnell_2013} found that both the \mmb\ and \msigma\ relations ``saturate'' towards the highest masses, underpredicting SMBH masses for the largest bright cluster galaxies. And an analysis by \citet{Sato-Polito_2024} finds that neither of the local relations can reproduce the high GWB amplitude. In either case, there is strong evidence to suggest that the local \mmb\ relation was different in the past in some way, and multimessenger studies such as this are an exiting route towards characterizing SMBH--galaxy coevolution.

\section{Summary and Conclusions}\label{sec:summary_and_conclusions}

In this work we implemented an evolving \mmb\ model which improves our ability to reproduce the GWB while maintaining consistency with astrophysically constrained models. We adapted the \textsc{holodeck} semi-analytic model to test for evolution in the \mmb\ relation and fit models to the GWB free-spectrum from \citet{Agazie_2023}. Our results show mild to strong preference for a positive evolution in the \mmb\ amplitude. When modeling the \mmb\ relation as $M_\mathrm{BH} = \alpha_0 (1 + z)^{\alpha_z} (M_{\mathrm{bulge}} / 10^{11}\ M_\odot)^{\beta_0}$ we find that $\alpha_z = \resalpha \pm \restd$.

We also studied the degeneracy between the GSMF and \mmb\ relation. The GWB requires massive SMBHs and we find that this population can be modeled with a top-heavy GSMF (which becomes a top-heavy BHMF via the local \mmb\ relation) or an increased \mmb\ amplitude (either via positive redshift evolution or a high local value). The \mmb\ relation remains the most influential component in GWB amplitude calculations. While an alternative GSMF offers an interesting solution, the GSMF has more robust observational constraints than the \mmb\ relation outside the local universe. Moreover, the GWB is most directly a probe of SMBH properties and so any inference about the GSMF is a secondary calculation. We therefore conclude that a SMBH-first growth model provides the best fits to the GWB while maintaining a high degree of consistency with EM-based observational constraints. 

Future work will investigate the effects of an evolving intrinsic scatter in the \mmb\ relation. It will also be important to test different evolutionary models using upcoming PTA data releases which will have higher signal-to-noise to better differentiate between models.


\begin{acknowledgments}

\rev{The authors would like to thank Joel Leja and CJ Harris for insightful discussions which aided and improved the interpretation of of this work. The authors additionally thank the anonymous referee for their thoughtful comments which led to an increase in the clarity of communication and analysis of our results.}

L.B.\ acknowledges support from the National Science Foundation under award AST-2307171 and from the National Aeronautics and Space Administration under award 80NSSC22K0808.
P.R.B.\ is supported by the Science and Technology Facilities Council, grant number ST/W000946/1.
S.B.\ gratefully acknowledges the support of a Sloan Fellowship, and the support of NSF under award \#1815664.
The work of R.B., R.C., X.S., J.T., and D.W.\ is partly supported by the George and Hannah Bolinger Memorial Fund in the College of Science at Oregon State University.
M.C., P.P., and S.R.T.\ acknowledge support from NSF AST-2007993.
M.C.\ was supported by the Vanderbilt Initiative in Data Intensive Astrophysics (VIDA) Fellowship.
Support for this work was provided by the NSF through the Grote Reber Fellowship Program administered by Associated Universities, Inc./National Radio Astronomy Observatory.
Pulsar research at UBC is supported by an NSERC Discovery Grant and by CIFAR.
K.C.\ is supported by a UBC Four Year Fellowship (6456).
M.E.D.\ acknowledges support from the Naval Research Laboratory by NASA under contract S-15633Y.
T.D.\ and M.T.L.\ received support by an NSF Astronomy and Astrophysics Grant (AAG) award number 2009468 during this work.
E.C.F.\ is supported by NASA under award number 80GSFC24M0006.
G.E.F., S.C.S., and S.J.V.\ are supported by NSF award PHY-2011772.
K.A.G.\ and S.R.T.\ acknowledge support from an NSF CAREER award \#2146016.
A.D.J.\ and M.V.\ acknowledge support from the Caltech and Jet Propulsion Laboratory President's and Director's Research and Development Fund.
A.D.J.\ acknowledges support from the Sloan Foundation.
N.La.\ acknowledges the support from Larry W. Martin and Joyce B. O'Neill Endowed Fellowship in the College of Science at Oregon State University.
Part of this research was carried out at the Jet Propulsion Laboratory, California Institute of Technology, under a contract with the National Aeronautics and Space Administration (80NM0018D0004).
D.R.L.\ and M.A.M.\ are supported by NSF \#1458952.
M.A.M.\ is supported by NSF \#2009425.
C.M.F.M.\ was supported in part by the National Science Foundation under Grants No.\ NSF PHY-1748958 and AST-2106552.
A.Mi.\ is supported by the Deutsche Forschungsgemeinschaft under Germany's Excellence Strategy - EXC 2121 Quantum Universe - 390833306.
The Dunlap Institute is funded by an endowment established by the David Dunlap family and the University of Toronto.
K.D.O.\ was supported in part by NSF Grant No.\ 2207267.
T.T.P.\ acknowledges support from the Extragalactic Astrophysics Research Group at E\"{o}tv\"{o}s Lor\'{a}nd University, funded by the E\"{o}tv\"{o}s Lor\'{a}nd Research Network (ELKH), which was used during the development of this research.
H.A.R.\ is supported by NSF Partnerships for Research and Education in Physics (PREP) award No.\ 2216793.
S.M.R.\ and I.H.S.\ are CIFAR Fellows.
Portions of this work performed at NRL were supported by ONR 6.1 basic research funding.
J.D.R.\ also acknowledges support from start-up funds from Texas Tech University.
J.S.\ is supported by an NSF Astronomy and Astrophysics Postdoctoral Fellowship under award AST-2202388, and acknowledges previous support by the NSF under award 1847938.
O.Y.\ is supported by the National Science Foundation Graduate Research Fellowship under Grant No.\ DGE-2139292.

Anishinaabeg gaa bi dinokiiwaad temigad manda Michigan Kichi Kinoomaagegamig. Mdaaswi nshwaaswaak shi mdaaswi shi niizhawaaswi gii-sababoonagak, Ojibweg, Odawaag, minwaa Bodwe’aadamiig wiiba gii-miigwenaa’aa maamoonjiniibina Kichi Kinoomaagegamigoong wi pii-gaa aanjibiigaadeg Kichi-Naakonigewinning, debendang manda aki, mampii Niisaajiwan, gewiinwaa niijaansiwaan ji kinoomaagaazinid.  Daapanaming ninda kidwinan, megwaa minwaa gaa bi aankoosejig zhinda akiing minwaa gii-miigwewaad Kichi-Kinoomaagegamigoong aanji-daapinanigaade minwaa mshkowenjigaade.

The University of Michigan is located on the traditional territory of the Anishinaabe people. In 1817, the Ojibwe, Odawa, and Bodewadami Nations made the largest single land transfer to the University of Michigan.  This was offered ceremonially as a gift through the Treaty at the Foot of the Rapids so that their children could be educated. Through these words of acknowledgment, their contemporary and ancestral ties to the land and their contributions to the University are renewed and reaffirmed.
\end{acknowledgments}

\begin{contribution}


C.M.\ was the primary lead for the formal analysis and investigation of this project. C.M.\ led the writing and editing of this manuscript and produced all figures and models. C.M.\, K.G.\, were responsible for the conceptualization and supervision of this work. C.M.\ and L.Z.K.\ developed and adapted the software and methodology applied here. K.G.\, L.Z.K.\,  J.S.\, and L.B.\ provided useful feedback and comments as well as aided in scientific interpretation of the results. Additional NANOGrav members are listed in alphabetical order, each contributed toward the collaboration-wide endeavor of pulsar timing  which culminated in evidence for the GWB. The work presented in this paper benefits from data collected and analyzed in the course of this search and would not be possible without this large-scale coordination.


\end{contribution}

%

\software{Astropy \citep{astropy:2013, astropy:2018, astropy:2022} -- SciPy \citep{Scipy_2020} -- NumPy \citep{numpy} -- Matplotlib \citep{matplotlib} -- kalepy \citep{kalepy} -- holodeck \citep{AgazieBHB_2023}, ceffyl \citep{Lamb_2023}}


\appendix

\section{GSMF Evolution Parameters and Scaling Relation Choice} \label{sec:app_gsmf}

\begin{figure*}[ht]\centering
    \includegraphics[width=\textwidth, keepaspectratio]{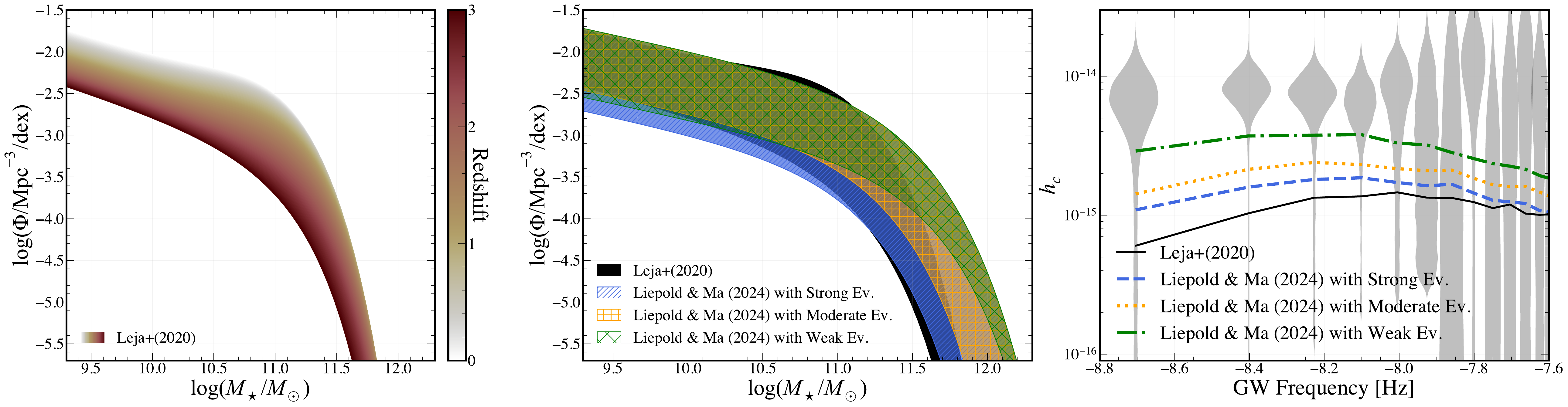}
    \caption{A comparison between the \citet{Leja_2020} GSMF (left red/blue and middle black) and our evolving versions of the \citet{Liepold_2024} GSMF (middle blue, orange, and green). The right panel shows the corresponding fiducial GWB spectra corresponding to each of these models. The GSMFs in the right and middle panels are all defined for $0 < z < 3$. The strongly evolving GSMF (blue) we consider is only negligibly different to that in \citet{Leja_2020} for $M_\star \geq 10^{11} M_\odot$ (which corresponds to the SMBH masses contributing to the GWB the most) for $ z \geq 1$. In this case we only see a minimal increase in amplitude to the GWB spectrum. In fact, even the weakly evolving GSMF (green) falls just short of the GWB data indicating that, a nearly non-evolving GSMF would be needed to match the GWB without changing other model parameters. For all GWB spectra shown here we use the \citet{Kormendy_Ho_2013} \mmb\ relation.}
    \label{fig:gsmf_options}
\end{figure*}

In this section, we describe the form of evolution we assume for the GSMF used in \liep\ and \liepev. We additionally explore the effects of alternative GSMFs not presented in the main text and briefly investigate the implications for the GWB and high-$z$ galaxy population. The different GSMFs and corresponding GWB spectra are shown in Figure \ref{fig:gsmf_options}. We also describe the effect of \mmb\ version on the predicted GWB spectrum at the end of this section and in Figure \ref{fig:kh_vs_mcma}.

The GSMF measured by \citet{Liepold_2024} uses only $z = 0$ galaxies and so they do not provide an explicit evolutionary form for the GSMF at higher redshift. For use in our analysis, we construct and test several models that bracket the plausible rates of GSMF evolution. A strongly evolving GSMF is predicted for $0 < z < 1$ \citep[e.g., ][]{De_Lucia_2007} though observations often show a lack of evolution in this redshift range \citep[e.g.,][]{Moustakas_2013, Bundy_2017, Leja_2020}. The \citet{Liepold_2024} GSMF offers observational support for this predicted evolution. The GSMF due to \citet{Leja_2020} does not show significant evolution for $0 < z < 1$ which, according to \citet{Liepold_2024} could be due to observational biases in the local universe. Large cosmological surveys do not have the appropriate local volume to catch the most massive galaxies and so the \citet{Liepold_2024} GSMF offers to complete the picture for the local high-mass GSMF. The \citet{Leja_2020} GSMF is consistent with many other studies and theory for $1 < z < 3$, and there is no strong foundation to believe that the GSMF in this redshift range is invalid. We can assume that the \citet{Leja_2020} GSMF is complete for $z \gtrsim 1$ and can therefore find an approximate evolution that matches the local \citet{Liepold_2024} GSMF and also the \citet{Leja_2020} for $z \gtrsim 1$.

Starting with the explicit $z = 0$ GSMF from \citet{Liepold_2024} and assuming a similar functional form to that in \citet{Leja_2020}, we tested three different forms of evolution represented by the shaded regions in Figure \ref{fig:gsmf_options}: (i) Weakly evolving (green): producing a GWB amplitude similar to that in \citet{Liepold_2024}, (ii) Moderately evolving (orange), and (iii) Strongly evolving (blue): Designed to match the high-mass end of the \citet{Leja_2020} GSMF by a redshift of $z \gtrsim 1$, all parameters for these models are given in Table \ref{tab:liep_strong_ev_priors}.

We believe the strongly evolving model to be the most faithful to the observational constraints from both \citet{Liepold_2024} and \citet{Leja_2020} hence this is the model we use for our analysis. The strongly evolving GSMF produces the lowest number densities at higher redshift compared to the other two evolutionary forms. The parameters for the strongly evolving GSMF were chosen such that the number density of massive galaxies is roughly similar to that of \citet{Leja_2020} by redshift $z \sim 1$. This option bridges the gap between the local GSMF from \citet{Liepold_2024} while maintaining consistency with \citet{Leja_2020} and therefore represents the solution that is in agreement with both observational constraints. This is the model we use in \liep\ and \liepev, the results of which are discussed in the main text in section \ref{sec:res_liep}.

The weakly evolving GSMF produces the highest GWB amplitude and is nearly consistent with the PTA data. In fact, only minor changes to, e.g., the hardening timescales would be needed to match the PTA data. This option represents the number density of galaxies that best reproduces the GWB starting from the local GSMF in \citet{Liepold_2024} and without significantly changing any other assumptions in our model. We only provide this approximate form and do not perform any fits because this GSMF is unrealistic and only used for demonstration purposes.

The moderately evolving GSMF represents a sort of ``compromise'' between the other two extremes. Interestingly, the posterior GSMF from \liep\ is very similar to this toy model we present here. The change in number density across $0 < z < 3$ is actually greater than in \citet{Leja_2020}, though it is positively offset at every redshift compared to their GSMF. This GSMF could be a realistic representation of the high-$z$ population, but would suggest that we are vastly underestimating galaxy stellar masses at these redshifts (e.g., via an incorrect initial mass function) or that cosmological surveys are largely incomplete (by a factor of 10 or more) in this regime. At this time we cannot evaluate the feasibility of this GSMF though, as emphasized earlier in section \ref{sec:discussion}, the inferences made for the GSMF from the GWB are to be treated with a high degree of caution.

\begin{deluxetable}{l ccc ccc ccc cc}
\tablewidth{0pt}

\tablehead{
\colhead{Evolution Type} & \colhead{$\phi_{*, 1,0}$} & \colhead{$\phi_{*, 1,1}$} & \colhead{$\phi_{*, 1,2}$} & \colhead{$\phi_{*, 2,0}$} & \colhead{$\phi_{*, 2,1}$} & \colhead{$\phi_{*, 2,2}$} & \colhead{$M_{\mathrm{c} ,0}$} & \colhead{$M_{\mathrm{c} ,1}$} & \colhead{$M_{\mathrm{c} ,2}$} & \colhead{$\alpha_1$} & \colhead{$\alpha_2$}
}
\startdata
Strong   & $-4.85$ & $-0.33$ & $-0.137$ & $-2.85$ & $-0.460$ & $0.057$ & $11.33$ & $0.0155$ & $-0.0413$ & $0.92$ & $-1.38$ \\
Moderate & $-4.85$ & $-0.26$ & $-0.110$ & $-2.85$ & $-0.370$ & $0.050$ & $11.33$ & $0.0200$ & $-0.0300$ & $0.92$ & $-1.38$ \\
Weak     & $-4.85$ & $-0.33$ & $-0.137$ & $-2.85$ & $-0.460$ & $0.057$ & $11.33$ & $0.1550$ & $-0.0413$ & $0.92$ & $-1.38$ \\
\enddata
\caption{The explicit evolution parameters for the strongly evolving version of the \citet{Liepold_2024} GSMF. Values under parameters with subscript 0 are equivalent to the combined $M_\star$ local values presented in \citet{Liepold_2024}. The other columns are the evolutionary parameters (see equations \ref{eqn:double_schechter}--\ref{eqn:mchar}).}
\label{tab:liep_strong_ev_priors}
\end{deluxetable}
\onecolumngrid

\section{Choice of Priors and Prior Tests} \label{sec:app_prior}

\rev{We chose astrophysically motivated priors as inputs to all our models. We investigated the impact that prior choice has on the models and found that our results are not sensitive to prior choice.}

\begin{figure}[ht]\centering
    \includegraphics[width=0.8\textwidth, keepaspectratio]{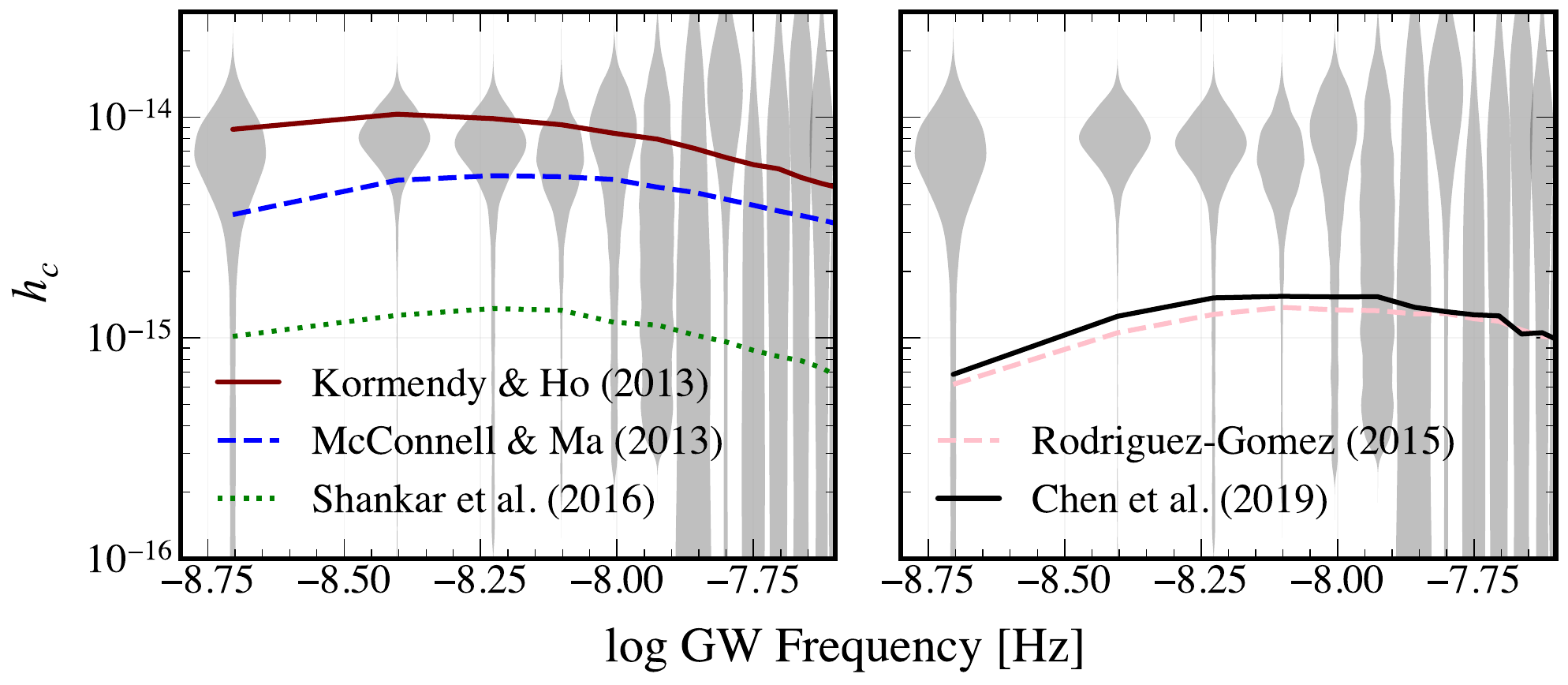}
    \caption{\rev{\textit{Left:} This figure demonstrates how the choice in scaling relation between \citet{Kormendy_Ho_2013}, \citet{McConnell_2013}, and \citet{Shankar_2016} affects the GWB spectrum. Because the \citet{Kormendy_Ho_2013} relation has a higher amplitude and steeper slope than \citet{McConnell_2013} and \citet{Shankar_2016}, the spectrum is broadly higher in amplitude with the biggest increase at the lower-frequency end.
    \textit{Right:} Here we demonstrate the impact of our choice in galaxy merger model on the GWB spectrum. The pair fraction prescription from \citet{Chen_2019} was used in the analysis performed by \citet{AgazieBHB_2023}, but here we use the merger rates from \citet{Rodriguez_Gomez_2015}. We feel this new prescription is a more astrophysically motivated model, but the impact on the GWB spectrum shape and amplitude is minimal.}}
    \label{fig:kh_vs_mcma}
\end{figure}

\rev{Several of our parameters, e.g., galaxy merger timescales and SMBH binary hardening timescales, do not have strong observational constraints. Instead, we make use of theoretically motivated prescriptions. The two studies we base our priors on both make use of the Illustris cosmological hydrodynamical simulations \citep{Vogelsberger_2014}. We use the galaxy merger rates due to \citet{Rodriguez_Gomez_2015} which provides an analytical prescription for merger rates that depends on galaxy mass and redshift. This is a different model to that used in \citet{AgazieBHB_2023} though we find that the impact on the resulting GWB spectrum is not significant. We ran a series of tests allowing only the galaxy merger rate parameters to vary and found that the posteriors always recovered the priors (and therefore the ``best fit'' model was identical to the prior model). We concluded that our models have minimal to no constraining power over these parameters hence they are fixed for \asm, \asmev, \asb, and \asbev.}

\rev{Numerous studies have investigated evolution of the physical separation of SMBH binaries, i.e., the ``hardening'' rate \citep[][Harris et al.\ in prep]{Begelman_1980, Quinlan_1996, Quinlan_1997, Yu_2002, Sesana_2015, Kelley_2017, Bhowmick_2024, Buttigieg_2025}. SMBH binary hardening rates are difficult to constrain observationally and so these studies rely on hydrodynamic and/or N-body simulations. Modeling the entire SMBH merger process is computationally challenging, however, leaving large uncertainties in hardening models \citep{Kelley_2017}. For this work, we use uniform priors on all free hardening parameters in our models. We see that the posterior distributions often push up against the limits of our prior range. In our testing, we found that the qualitative results of this work were not sensitive to the prior ranges we assume for these parameters. For example the hardening timescale, $\tau_f$ lower limit is 0 Gyr and cannot go lower despite the posterior distribution pushing against this lower limit. Additionally the $\nu_{\mathrm{inner}}$ parameter lower limit is roughly at the limit at which the timescale of the stellar scattering regime is minimized and lower values of $\nu_{\mathrm{inner}}$ have a negligible impact on the GWB spectrum. Finally the third hardening parameter we sample is the characteristic radius, $R_{\mathrm{char}}$, which indicates the point at which the dominant hardening mechanism transitions from dynamical friction to stellar scattering. This radius corresponds to the sphere of influence of the SMBHs and is on the order of 10s of pc. Our priors range from 2--20 pc, which is physically motivated \citep{Kelley_2017}. Allowing for larger values of $R_{\mathrm{char}}$ can be useful for studies focused on constraints at small binary separations, but these parameters are not the focus of this study so we chose a realistic, but uniform range of values. We also note that there is a degeneracy between $R_{\mathrm{char}}$ and $\nu_{\mathrm{inner}}$ such that large values of $R_{\mathrm{char}}$ correspond to larger values of $\nu_{\mathrm{inner}}$. Therefore changing the prior range on $R_{\mathrm{char}}$ may affect the posterior values for, e.g., $\nu_{\mathrm{inner}}$, but it does not notably impact the results of GSMF or \mmb\ parameters. Improving the hardening parameter prior choice for GWB studies requires a dedicated multimessenger study. This is outside the scope of this work, but will be included in future investigations (e.g., Harris et al.\ in prep).}

\begin{figure*}[ht]\centering
    \includegraphics[width=\textwidth, keepaspectratio]{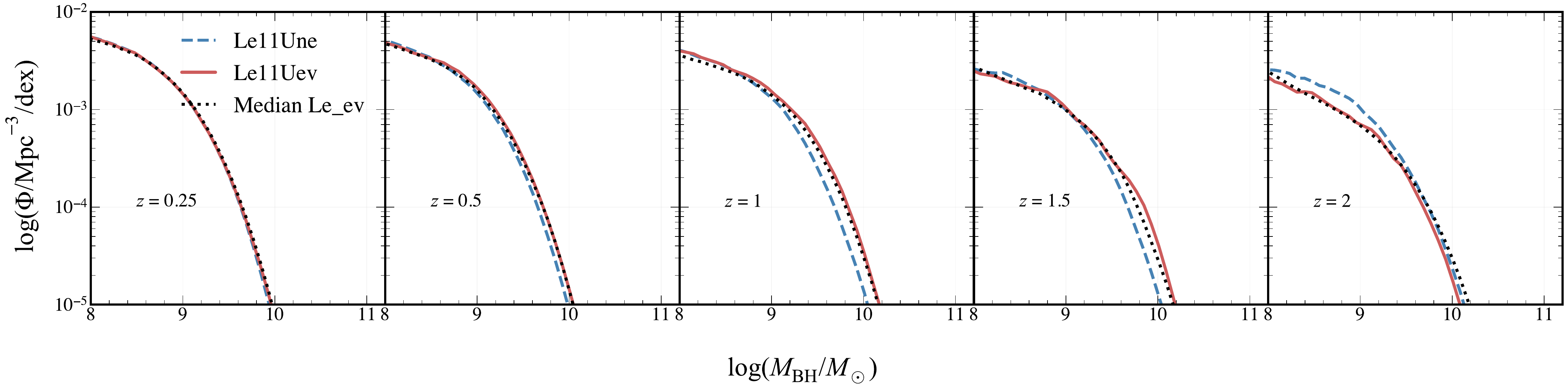}
    \caption{\rev{The posterior BHMFs associated with our two models with uniform GSMF priors. The model associated with the blue dashed line sampled only the 11 GSMF parameters while the red solid line model additionally sampled the $\alpha_z$ parameter. We additionally show the posterior BHMF from Le\_ne as a black dotted line. This is the same as the red solid line in Figure \ref{fig:bhmf}. We omit error bars for clarity though they are slightly larger than those in Figure \ref{fig:bhmf}. There is broad agreement in SMBH number density across all redshift bins between the three models shown here. This demonstrates that our models have constraining power over the BHMF and are not prior dominated.}}
    \label{fig:bhmf_uniform}
\end{figure*}

\rev{For the GSMF and \mmb\ relation, we used models based on electromagnetic observations \citep{Leja_2020, Kormendy_Ho_2013, Liepold_2024}. To evaluate the constraining power of our models over the GSMF, we ran a model sampling only the 11 GSMF parameters with uniform priors (Le11Une) and repeated this model with a free $\alpha_z$ parameter (Le11Uev). We found that the results of this test were consistent with the results of our study. The posterior distributions deviated from the uniform priors and produced BHMFs that are consistent with each other and also the BHMFs from our models with informed priors. The BHMFs from these models are shown in Figure \ref{fig:bhmf_uniform}, we additionally over-plot the median BHMF from the Le\_ne models from Figure \ref{fig:bhmf}. We see that all three BHMFs return remarkably similar number densities in all redshift bins though Le11Une falls slightly short of the two evolving \mmb\ models, a pattern also present in Figure \ref{fig:bhmf}. The posterior \mmb\ evolution parameter from Le11Uev is $\alpha_z = 0.98 \pm 1.1$ and the distribution is 82.8\% positive. While this is not statistically significant evolution, we believe this result supports our overall conclusions and is a good indicator that our models are not prior dominated for the GSMF parameters.}

\rev{Similar to how we tested different options for the GSMF, we additionally considered the effect of the version of the \mmb\ relation we use. In Figure \ref{fig:kh_vs_mcma} we demonstrate the impact of scaling relation choice on the GWB spectrum. The version of the scaling relation from \citet{Kormendy_Ho_2013} generally produces bigger SMBHs (and therefore higher GWB amplitudes) than other options \citep[see, e.g., ][]{Simon_2023}. \citet{Liepold_2024} use the \citet{McConnell_2013} version of the \mmb\ relation, which generally produces lower GWB amplitudes compared to \citet{Kormendy_Ho_2013}. We additionally show the GWB spectrum using the scaling relation from \citet{Shankar_2016}. This comparison demonstrates that, wherever our spectra fall short of the GWB data, using a relation with a lower amplitude would only exacerbate this problem. This also has implications for our choices for the evolution we approximate for the \citet{Liepold_2024} GSMF.}

\section{Reliability of Methods} \label{sec:app_comp}

\begin{figure*}[ht]\centering
    \includegraphics[width=\textwidth, keepaspectratio]{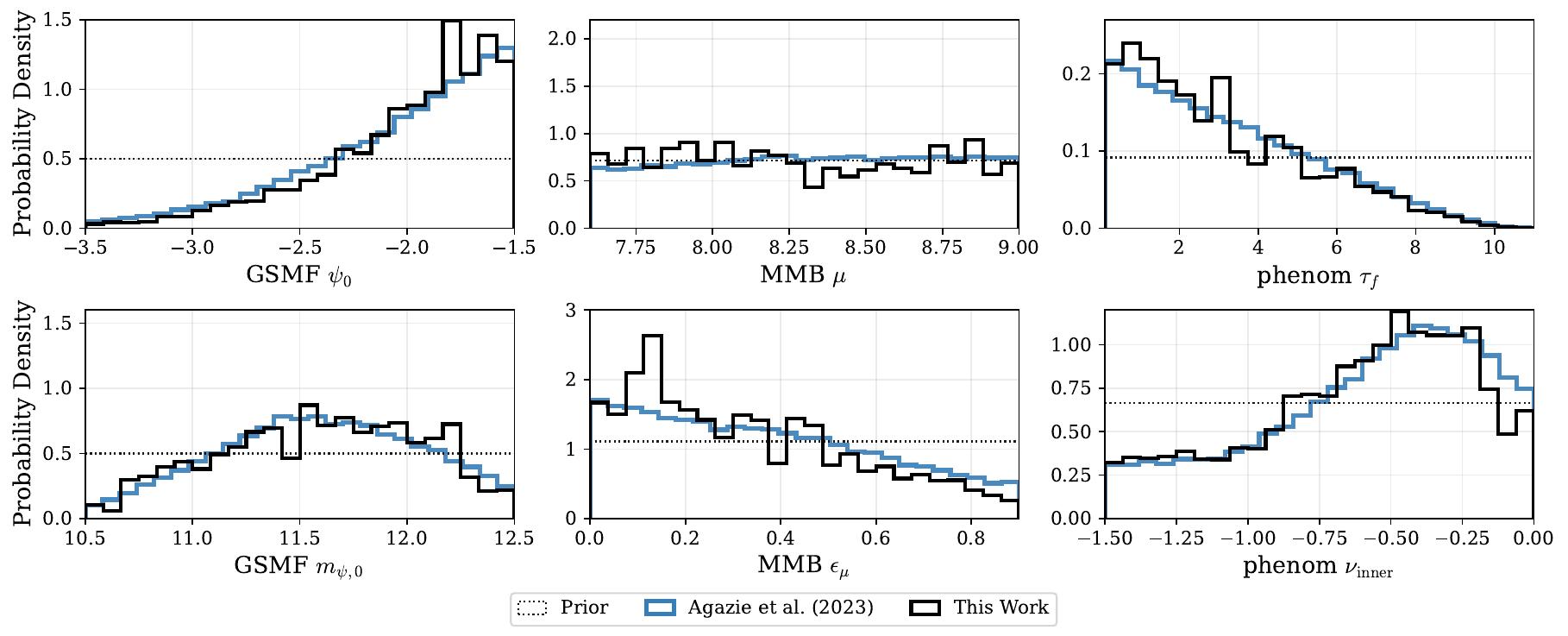}
    \caption{Posterior distributions for the same model set up from \citet{AgazieBHB_2023} when using our method (blue) versus their (black). We find that our results are in good agreement with theirs when using the same model.}
    \label{fig:comp_15}
\end{figure*}

Here we demonstrate that our ``direct-likelihood" method reproduces the Gaussian process-interpolated MCMC-like methods of \citet{AgazieBHB_2023}. In Figure \ref{fig:comp_15} we show a comparison between the posterior distributions from \citet{Agazie_2023} compared to our results when using the same prior model set up. We see that the two methods return negligibly different posterior distributions. The method we use in this work is also discussed in more depth in Appendix C of \citet{AgazieBHB_2023}.

In this work we discovered a minor limitation of the direct-likelihood method which impacted the \asb\ model. It is more difficult for the posterior distributions to deviate from the priors because extreme values (e.g. more than 3$\sigma$ above or below the prior median) have a low chance of being randomly sampled, regardless of how well that value may describe the data. We see this limitation in our model with the lowest number of sampled parameters, \asb. The best-fit GWB spectrum from this model returned a notably low log-likelihood compared to the other models \revii{which means it} produces a worse overall fit (see Figures \ref{fig:all_hcvsf_comp} and \ref{fig:xi_vs_alpha}). We can see, from \asmev\ and \asbev, that a value of $\alpha(z) > 9.0$ is needed to reproduce the GWB amplitude without increasing galaxy number density. If the \mmb\ relation is not allowed to evolve, the prior for $\alpha_0$ would need to be sampled at a value $\gtrsim 6\sigma$ away from the prior mean (assuming a normal distribution where $\alpha_0 = 8.69 \pm 0.055$), and therefore extremely unlikely / impossible to be sampled. Without interpolating, any value that is not sampled simply does not have a corresponding GWB spectrum and the best-fit parameters may therefore not describe the data. This effect can be mitigated by choice of priors, but prior choice has its own impact on models and so future studies should consider the impact of this effect when using this method.

\revii{It is important to emphasize that the likelihoods we calculate are a measure of how well a given best-fit spectrum describes the observed GWB spectrum, but they are not marginalized over the respective parameter spaces. This means that the likelihoods to not contain information about prior-posterior agreement/disagreement and therefore do not indicate goodness of fit to EM data. These likelihoods cannot be used alone to evaluate which models are better models of \textit{both} the GWB and EM data. To evaluate consistency with EM data we make use of other quantities such as the Kullback-Leilbler divergence, $D_{KL}$, and GSMF boost, $\Xi$. The ``best'' models have high likelihoods (good fits to the GWB) in addition to low $D_{KL}$ and $\Xi$ values (good agreement with the EM data).}

\section{Corner Plots} \label{sec:app_corner}

The full corner plots for each of our \modelnum\ models. Note that the variables pertaining to galaxy merger rates include ``GMR'' in the subscript to distinguish from other parameters which use similar variables. For the parameters in the model not detailed in this paper we refer the reader to \citet{AgazieBHB_2023} (and references therein) for a complete discussion. In each corner plot, the blue histograms / contours represent the posterior distributions. The priors are shown by the gray histograms.

\begin{figure*}[ht]\centering
    \includegraphics[width=\textwidth, keepaspectratio]{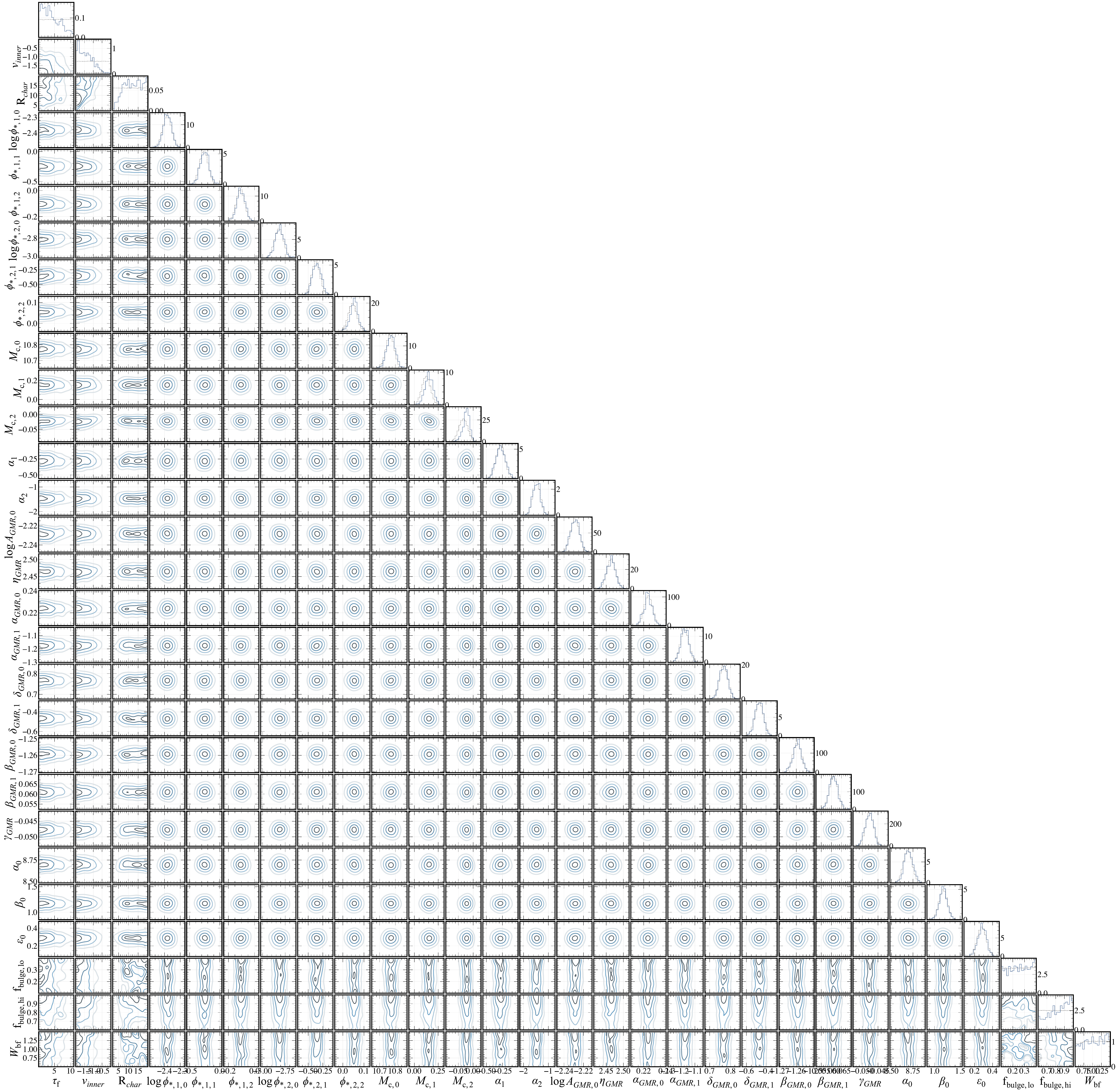}
    \caption{The complete corner plot for \asa. The complete figure set (eight images) is available in the online journal.}
    \label{fig:cornerplots}
\end{figure*}

\figsetstart
\figsetnum{1}
\figsettitle{Model Output Corner Plots}
\figsetgrpstart
\figsetgrpnum{1}
\figsetgrptitle{Complete corner plot for \asa.}
\figsetplot{asa_corner.pdf}
\figsetgrpnote{The full corner plots for each of our \modelnum\ models.}
\figsetgrpend
\figsetgrpnum{2}
\figsetgrptitle{Complete corner plot for \asaev.}
\figsetplot{asa_ev_corner.pdf}
\figsetgrpnote{The full corner plots for each of our \modelnum\ models.}
\figsetgrpend
\figsetgrpnum{3}
\figsetgrptitle{Complete corner plot for \liep.}
\figsetplot{lie_corner.pdf}
\figsetgrpnote{The full corner plots for each of our \modelnum\ models.}
\figsetgrpend
\figsetgrpnum{4}
\figsetgrptitle{Complete corner plot for \liepev.}
\figsetplot{lie_ev_corner.pdf}
\figsetgrpnote{The full corner plots for each of our \modelnum\ models.}
\figsetgrpend
\figsetgrpnum{5}
\figsetgrptitle{Complete corner plot for \asm.}
\figsetplot{asm_corner.pdf}
\figsetgrpnote{The full corner plots for each of our \modelnum\ models.}
\figsetgrpend
\figsetgrpnum{6}
\figsetgrptitle{Complete corner plot for \asmev.}
\figsetplot{asm_ev_corner.pdf}
\figsetgrpnote{The full corner plots for each of our \modelnum\ models.}
\figsetgrpend
\figsetgrpnum{7}
\figsetgrptitle{Complete corner plot for \asb.}
\figsetplot{asb_corner.pdf}
\figsetgrpnote{The full corner plots for each of our \modelnum\ models.}
\figsetgrpend
\figsetgrpnum{8}
\figsetgrptitle{Complete corner plot for \asbev.}
\figsetplot{asb_ev_corner.pdf}
\figsetgrpnote{The full corner plots for each of our \modelnum\ models.}
\figsetgrpend
\figsetend

\bibliography{ref}{}
\bibliographystyle{aasjournalv7}



\end{document}